\documentclass[aps,prc,preprint,tightenlines,superscriptaddress,showpacs,%
amssymb,byrevtex,nofootinbib]{revtex4}
\usepackage[dvips]{graphicx,color}
\usepackage{amsmath}
\usepackage{mathrsfs}
\usepackage{dcolumn}
\usepackage{epsfig}
\usepackage{bm}
\usepackage{xcolor}
\usepackage{ulem}

\begin{document}
\title{Nucleon resonance structure in the finite volume of lattice QCD}
\author{Jia-Jun Wu}
\affiliation{Special Research Center for the Subatomic Structure of Matter (CSSM),
             Department of Physics, University of Adelaide Adelaide 5005, Australia}
\author{H. Kamano}
 \affiliation{KEK Theory Center, Institute of Particle and Nuclear Studies (IPNS),
High Energy Accelerator Research Organization (KEK), Tsukuba, Ibaraki 305-0801, Japan}
\affiliation{J-PARC Branch, KEK Theory Center, IPNS, KEK, Tokai, Ibaraki 319-1106, Japan}
\author{T.-S. H. Lee}
\affiliation{Physics Division, Argonne National Laboratory, Argonne, Illinois 60439, USA}
\author{D. B. Leinweber}
\affiliation{Special Research Center for the Subatomic Structure of Matter (CSSM),
             Department of Physics, University of Adelaide Adelaide 5005, Australia}
\author{A.W. Thomas}
\affiliation{Special Research Center for the Subatomic Structure of Matter (CSSM),
             Department of Physics, University of Adelaide Adelaide 5005, Australia}

\begin{abstract}
An approach for relating the nucleon resonances extracted from $\pi N$ reaction data to lattice QCD
calculations has been developed by using the finite-volume Hamiltonian method.  Within models of
$\pi N$ reactions, bare states are introduced to parametrize the intrinsic excitations of the
nucleon.  We show that the resonance pole positions can be related to the probability $P_{N^*}(E)$
of finding the bare state, $N^*$, in the $\pi N$ scattering states in infinite volume.  We further
demonstrate that the probability $P_{N^*}^V(E)$ of finding the same bare states in the
eigenfunctions of the underlying Hamiltonian in finite volume approaches $P_{N^*}(E)$ as the volume
increases.  Our findings suggest that the comparison of $P_{N^*}(E)$ and $P_{N^*}^V(E)$ can be used
to examine whether the nucleon resonances extracted from the $\pi N$ reaction data within the
dynamical models are consistent with lattice QCD calculation.
We also discuss the measurement of $P_{N^*}^V(E)$ directly from lattice QCD.
The practical differences between our approach and the approach
using the L\"uscher formalism to relate LQCD calculations to the nucleon resonance poles embedded
in the data are also discussed.

\end{abstract}

\maketitle

\section{Introduction}

One of the most challenging problems in modern hadron physics is to understand the spectra of
baryons and mesons within Quantum Chromodynamics (QCD); the fundamental theory of the strong
interactions. It is therefore important to investigate how the properties of the excited nucleons
can be understood using lattice QCD calculations (LQCD). Important progress in this direction has
been made in recent years  \cite{lqcd-1,lqcd-2} and the accuracy of the results is expected to
improve rapidly in the near future. It is therefore necessary to address the question of how LQCD
results can be related to the experimental data and, further, how they may be used to understand
the manner in which nucleon excited states emerge from non-perturbative QCD.

The excited nucleons are unstable and coupled with the meson-nucleon continuum to form nucleon
resonances ($N^*$). Thus the properties of excited nucleons can only be studied by analyzing the
nucleon resonances extracted from data, such as meson production reactions induced by pions,
photons and electrons.  Accordingly, it is necessary to develop an approach to relate the resonance
parameters, which are defined on the complex energy ($E$)-plane, to the results from LQCD
calculations.
Several different approaches have been developed.


The first approach \cite{jlab-lqcd-1,jlab-lqcd-2} is to use the L\"uscher formalism \cite{luch-1,luch-2,luch-3} to extract the scattering phase shifts from the spectrum calculated   using LQCD.  A K-matrix model, with an appropriate phenomenological parametrization, is then constructed to fit the extracted phase shifts.  The resonance parameters extracted within the constructed K-matrix model are then compared with those listed by Particle Data Group \cite{pdg}(PDG) .

The second approach is the finite-volume Hamiltonian method (FVH method) developed in Refs.~\cite{ad-1, ad-2, ad-3, ad-4, ad-5, ad-6}. This starts with the construction of a Hamiltonian to fit the data of the processes under consideration.  The resulting Hamiltonian is then used to predict the spectrum in finite volume and that is compared with the spectrum calculated from LQCD. Agreement between these spectra implies that the LQCD calculation gives the same resonance parameters embedded in the data through the constructed Hamiltonian. Alternatively, this second approach can also be used to fit the LQCD energy spectrum and the
resulting Hamiltonian then used to calculate phase shifts for comparison with experimental data.

The approach developed in Refs.~\cite{bonn-lqcd-1,bonn-lqcd-2,bonn-lqcd-3} also involves a formulation of the problem within a finite-volume, starting with the scattering equations deduced from unitarized chiral perturbation theory.  It has also been used to extract resonance parameters by an appropriate analytic continuation.

If the spectrum calculated from LQCD is of very high accuracy and covers a sufficiently wide energy
region, within which the experimental data for investigating a particular nucleon resonance are
also accurate and complete (as reviewed in Ref.~\cite{shkl}),
then the first and second approaches are equally
valid.  This is supported by the results from a study \cite{lwk} of resonance extractions and the
FVH method.  It was demonstrated within several exactly soluble models that the extracted resonance
parameters are independent of the model used in the resonance extraction as far as the partial-wave
amplitude data within the sufficiently wide region near the considered resonance are fitted
$precisely$ (i.e., within $1\%$ considered in Ref.~\cite{lwk}).

Unfortunately, this ideal situation does not exist in reality for investigating nucleon resonances
at the present time.  The scattering amplitudes determined from either the experimental data or the
LQCD spectrum and L\"uscher's formula have intrinsic errors associated with the unavoidable
systematic and statistical errors. Thus the extracted resonance parameters, widths and residues, can
depend significantly on the parametrization of the K-matrix and the form of the Hamiltonian used
to fit the determined scattering amplitudes within the errors, particularly in the higher-mass
region as observed in Ref.~\cite{anl-osaka}.

The purpose of the present work is to apply the FVH method in the development of an approach to
relate the nucleon resonances to LQCD calculations.  Instead of the separable potential models used
in the previous FVH studies \cite{ad-1,ad-2,ad-3,ad-4}, dynamical $\pi N$ reaction models based on
meson-exchange mechanisms are used.  We will start with a one-channel dynamical model (the Sato-Lee
(SL) model) developed in Ref.~\cite{sl}.  This model, with one bare state in the $P_{33}$
partial-wave, is consistent with the well-accepted interpretation
\cite{Theberge:1980ye,Thomas:1982kv} that the $\Delta$ (1232) resonance is made of a quark core and
a meson cloud.

We first apply the SL Hamiltonian to confirm the results, as established in
\cite{ad-1,ad-2,ad-3,ad-4,ad-5, ad-6}, that the FVH method is equivalent to using L\"uscher's formalism in
relating the spectrum in finite volume to the scattering amplitudes in infinite volume. We then
observe that the probability $P_\Delta(E)$ of finding the bare $\Delta$ state in the $\pi N$
scattering wave function contains resonance information which can be verified on the real-$E$ axis,
which is in turn accessible to experiments.  We then demonstrate that an energy-averaged
probability $P_\Delta^V(E, L)$ of finding the bare $\Delta$ in the eigenstate of the
Hamiltonian in finite volume approaches $P_\Delta(E)$ as the volume size $L$ increases.
This result indicates that $P_\Delta^V(E, L)$ from LQCD calculations can be related
directly to the nucleon resonance information extracted within the given dynamical model.  Clearly,
this is rather different from Approach 1 mentioned above, which uses the K-matrix model to extract
nucleon resonance properties from the spectrum obtained in LQCD calculations.

We next consider a three-channel, meson-exchange $\pi N$ model within which there is one bare state
in each partial-wave.  The parameters of this model are determined by fitting the empirical $s$ and
$p$ partial-wave amplitudes up to 1.6 GeV.  This allows us to examine the more complex situation in
which two resonances are associated with the same bare state in the $P_{11}$ partial wave.  This is
similar to the results obtained from the analysis of Ref.~\cite{anl-osaka}.  Here we examine
closely the differences between using the FVH method and L\"uscher's formalism to relate the
multi-channel scattering amplitudes and the associated nucleon resonances to the LQCD calculations
through the spectrum in finite volume. We then demonstrate that for the multi-channel case the
probability $P_\Delta^V(E, L)$ in finite volume also approaches the probability
$P_\Delta(E)$ in infinite volume, as the volume increases.

Our findings suggest that the comparison of $P_{N^*}(E)$ and $P_{N^*}^V(E, L)$ can be
used to examine whether the nucleon resonances extracted from the $\pi N$ reaction data within the
dynamical models are consistent with LQCD. We will discuss possible LQCD calculations of
$P_{N^*}^V(E, L)$ for interpreting the bare states of the dynamical models.  This
provides a new method to extract the properties of hadrons directly from LQCD calculations through
measuring $P_{N^*}^{V}(E, L)$.  We anticipate the formalism developed herein will be
applied in next-generation lattice QCD calculations extracting the complete spectrum through the
incorporation of non-local meson-baryon interpolating fields.

In section II, we present details of the calculations based on a dynamical Hamiltonian model in
infinite volume and in a finite volume.  The results for the SL model and the three-channel model
are presented in sections III and IV, respectively.  In section V and Appendix A, we discuss
possible LQCD calculations of $P_{N^*}^V(E, L)$.  A summary and some discussion of
possible future directions are given in section VI.

\section{Scattering solutions from Dynamical Hamiltonians}

The Hamiltonian of the dynamical model we will consider is defined by
\begin{eqnarray}
H=H_0+ H_I \, ,
\label{eq:dm-h}
\end{eqnarray}
where $H_0$ is the free Hamiltonian. The interaction Hamiltonian is taken to have the following
form
\begin{eqnarray}
H_I= \sum_{i=1,n_c} g_{N^*,i} + \sum_{i,j=1,n_c}v_{i,j} \, ,
\label{eq:dm-hi}
\end{eqnarray}
where $n_c$ is the number of meson-baryon channels considered, $g_{N^*,i}$ is the vertex
interaction defining the decay of a bare $N^*$ state into the $i-$th meson-baryon channel and
$v_{i,j}$ is the two-body meson-baryon interaction between channels $i$ and $j$.  In both the SL
model and the three-channel model, the interactions $v_{i,j}$ are calculated from meson-exchange
mechanisms derived from phenomenological Lagrangians.

In the following two subsections, we write down the formulas required to calculate the scattering
amplitudes from the Hamiltonian Eqs.~(\ref{eq:dm-h}) and ({\ref{eq:dm-hi}) in infinite volume as well as
  in finite volume.

\subsection{Solutions in infinite volume}
Based on the Hamiltonian defined by Eqs.~(\ref{eq:dm-h}) and (\ref{eq:dm-hi}), it is known
\cite{sl,anl-osaka} that the scattering amplitudes of each partial-wave can be written as
\begin{eqnarray}
T_{i,j}(k,k';E) = t^{bg}_{i,j}(k,k';E) + t^{\rm res}_{i,j}(k,k';E) \, .
\label{eq:t-tot}
\end{eqnarray}
Here and in the rest of this paper the indices $(i,j)$ also specify the quantum numbers associated
with the meson-baryon channel, namely, the orbital angular momentum ($L$), total spin ($S$), total
angular momentum ($J$), parity ($P$), and isospin ($I$).  The 'background' amplitudes
$t^{bg}_{i,j}(k,k',E)$ are calculated from the meson-baryon interactions by
\begin{eqnarray}
t^{bg}_{i,j}(k,k';E) &=&v_{i,j}(k,k';E)\nonumber \\
&&+ \sum_{m}\int k^{''\,2} dk^{''}
v_{i,m}(k^{'},k^{''};E)\,\frac{1}{E-E_{M_m}(k^{''})-E_{B_m}(k^{''})+i\epsilon}
t^{bg}_{m,j}(k^{''},k';E) \, . \nonumber \\
&&
\end{eqnarray}
The resonant amplitudes are
\begin{eqnarray}
t^{\rm res}_{i,j}(k,k';E) =\frac{\bar{\Gamma}^\dagger_i(k;E)\,\bar{\Gamma}_j(k';E)}
{E-m_0-\Sigma(E)}\, ,
\label{eq:t-res}
\end{eqnarray}
where the dressed vertex functions are
\begin{eqnarray}
\bar{\Gamma}^\dagger_i(k;E) &=& \Gamma^\dagger_{N^*,i}(k) + \sum_{m}\int k^{'\,2} dk^{'}
t^{bg}_{i,m}(k,k^{'};E)\,\frac{1}{E-E_{M_m}(k^{'})-E_{B_m}(k^{'})+i\epsilon}
\Gamma^\dagger_{N^*,m}(k^{'}) \, , \nonumber \\
&& \\
\bar{\Gamma}_j(k;E) &=& \Gamma_{N^*,j}(k) + \sum_{m}\int k^{'\,2} dk^{'}
\Gamma_{N^*,m}(k^{'})\frac{1}{E-E_{M_m}(k^{'})-E_{B_m}(k^{'})+i\epsilon}
t^{bg}_{m,j}(k^{'},k;E)\, , \nonumber \\
&&
\end{eqnarray}
and the self-energy of the $N^*$ is
\begin{eqnarray}
\Sigma(E) = \sum_{m}\int k^{'\,2} dk^{'}
\Gamma_{m}(k^{'})\frac{1}{E-E_{M_m}(k^{'})-E_{B_m}(k^{'})+i\epsilon}\bar{\Gamma}^\dagger_m(k^{'};E)
\, .
\label{eq:selfe}
\end{eqnarray}
As developed in Refs.~\cite{ssl-1,sjklt}, the resonance poles $E_{res}$ of the scattering
amplitudes $T_{i,j}$ can be found from the resonant part $t^{\rm res}_{i,j}$ of Eq.~(\ref{eq:t-tot}).
{}From the expression Eq.~(\ref{eq:t-res}), it is clear that $E_{res}$ can be obtained by solving
the following equation on the complex$-E$ plane
\begin{eqnarray}
E_{res}-m_0 - \Sigma(E_{res})=0\, .
\label{eq:pole}
\end{eqnarray}
This equation can lead to many poles.  However, only the poles near the physical region are
relevant to the physical observables.  The energies of these resonance poles in general have the
form $E_{res} = E_R - i E_I$ with $E_R, E_I > 0$.  In the Argonne National Laboratory-Osaka
University (ANL-Osaka) analysis \cite{anl-osaka}, only those poles with $E_I < 200$ MeV are
considered to be related to excited nucleon states through their coupling with the meson-baryon
continuum.

We next use $t^{\rm res}_{\pi N,\pi N}$ of the total amplitude $T$ of Eq.~(\ref{eq:t-tot}) to define
the resonant cross section of $\pi N$ elastic scattering as
\begin{eqnarray}
\sigma^{\rm res}(E)&=& \frac{(4\pi)^2}{k_{\pi N}^2}\rho^2_{\pi N}(E)\,\frac{2J+1}{2}
\left |t^{\rm res}_{\pi N,\pi N}(k_{\pi N},k_{\pi N},E)\,\right |^2\, , \nonumber \\
&=&\frac{(4\pi)^2}{k_{\pi N}^2}\rho^2_{\pi N}(E)\,\frac{2J+1}{2}
\left |\frac{\bar{\Gamma}^\dagger_{\pi N}(k_{\pi N};E)\,\bar{\Gamma}_{\pi N}(k_{\pi N};E)}
{E-m_0-\Sigma(E)}\right |^2 \, ,
\label{eq:sigma-r0}
\end{eqnarray}
where $k_{\pi N}$ is the $\pi N$ on-shell momentum, and $\rho_{\pi N}(E) = \pi\, k_{\pi N}\,
E_N(k)\, E_\pi(k)\,/E$.  We can cast Eq.~(\ref{eq:sigma-r0}) into the following form
\begin{eqnarray}
\sigma^{\rm res}(E)&=&
\left |\frac{1}{E-m_0-\Sigma(E)}\right |^2 \left[\,\frac{(4\pi)^2}{k_{\pi N}^2}\rho^2_{\pi N}(E)\,\frac{2J+1}{2}
\left |\bar{\Gamma}_{\pi N}(k_{\pi N},;E)\,\right |^4 \,\right] \, .
\label{eq:sigma-r}
\end{eqnarray}
Because of the condition Eq.~(\ref{eq:pole}), one can consider that $\sigma^{\rm res}_{\pi N}(E)$
contains the resonance information on the real-$E$ axis which is accessible to experiments.  In
some cases it is possible to cast the expression Eq.~(\ref{eq:sigma-r}) into the Breit-Wigner form
in the region where $(E_R-2E_I) \leq E \leq (E_R + 2E_I)$. But the parameters of
the resulting Breit-Wigner resonances will differ from those of the extracted resonance poles,
which are known \cite{ssl-1,bohm} to be the energies of the eigenstates of the underlying
Hamiltonian with outgoing wave boundary condition.

We now introduce a quantity which can be related to $\sigma^{\rm res}$ and which can also be defined
within the finite-volume formulation.  We start by examining the scattering wave function with an
incident plane-wave state in the $i=1=\pi N$ channel. It is defined by the total amplitude
Eq.~(\ref{eq:t-tot}) :
\begin{eqnarray}
|\Psi^{(+)}_{E,\pi N}\rangle=\left [1+\frac{1}{E-H_0+i\epsilon}T(E)\,\right ] |k_{\pi N}\rangle\,\,,
\label{eq:pin-wf}
\end{eqnarray}
where $|k_{\pi N}\rangle$ is the incoming $\pi N$ plane-wave state.  It is well known from standard
reaction theory \cite{gw} that
\begin{eqnarray}
(H_0+H_I)\, | \Psi^{(+)}_{E,\pi N} \rangle = E \, |\Psi^{(+)}_{E,\pi N}\rangle \, .
\label{eq:psi-inf}
\end{eqnarray}
We can use the definition Eq.~(\ref{eq:pin-wf}) and the solutions given by
Eqs.~(\ref{eq:t-tot})-(\ref{eq:selfe}), to verify Eq.~(\ref{eq:psi-inf}) explicitly and also to
obtain the following relation
\begin{eqnarray}
\langle N^*|\Psi^{(+)}_{E,\pi N}\rangle = \frac{\bar{\Gamma}_{\pi N}(k_{\pi N};E)}{E-m_0 -
  \Sigma(E)} \, .
\label{eq:n-wf}
\end{eqnarray}
Thus the probability of finding the bare $N^*$ state in the $\pi N$ scattering wave function is
\begin{eqnarray}
p_{\pi N}(E)&=&\left |\langle N^*|\Psi^{(+)}_{E,\pi N} \rangle \right |^2
=\left |\frac{\bar{\Gamma}(k_{\pi N};E)}{E-m_0-\Sigma(k)}\right |^2 \, .
\label{eq:prob-pin-ifv}
\end{eqnarray}
By comparing $p_{\pi N}(E)$ and $\sigma^{\rm res}(E)$ (Eq.~(\ref{eq:sigma-r})), we can see that $p_{\pi
  N}(E)$ contains the resonance information on the real-$E$ axis which is accessible to
experiments.

One can generalize the above formula to define $p_i(E)$ for any channel $i=1, ..., n_c$ included in the
model.  We define the total probability of finding the bare $N^*$ state in the scattering wave
function as
\begin{eqnarray}
P_{N^*}(E)=\frac{1}{Z}[\sum_{i=1,n_c} \rho_i(E)\,p_{i}(E)\,]\, ,
\label{eq:pe-all}
\end{eqnarray}
with
\begin{eqnarray}
\rho_i(E)=\pi\, k_i\,E_{M_i}(k_i)\, E_{B_i}(k_i) \, ,
\end{eqnarray}
where $k_i$ is the on-shell momentum of channel $i$, and
\begin{eqnarray}
p_{i}(E)&=&\left |\langle N^*|\Psi^{(+)}_{E,i} \rangle \right |^2 \, ,\label{eq:prob-i} \\
&=& \left |\frac{\bar{\Gamma}(k_{i};E)}{E-m_0-\Sigma(E)}\right |^2 \, ,
\label{eq:prob-ii} \\
Z&=&\sum_{i=1,n_c}\int_{E_{{th}_i}}^\infty\,dE\, \rho_i(E)\,p_{i}(E) \, .
\end{eqnarray}
Here $E_{{th}_i}$ is the threshold energy in the $i$-th channel.
Clearly, we can write
\begin{eqnarray}
P_{N^*}(E)=\left |\frac{1}{E-m_0-\Sigma(E)}\right |^2\frac{1}{Z}\sum_{i}\rho_i(E)\, \left |
\bar{\Gamma}_i(k_i,E)\,\right |^2 \, .
\label{eq:pe-tot}
\end{eqnarray}
By comparing Eqs.~(\ref{eq:sigma-r}) and (\ref{eq:pe-tot}), we observe that $P_{N^*}(E)$ has a similar
energy-dependence to $\sigma^{\rm res}(E)$ and that it also contains the resonance information on the
real-$E$ axis which is accessible to experiments.

\subsection{Solution in a finite volume}
In a periodic volume characterized by side length $L$, the quantized three momenta of mesons and
baryons must be $k_n = \sqrt{n}\frac{2\pi}{L}$ for integers $n = 0,1,2,\ldots$. Because of the
presence of a bare $N^*$ state in the dynamical Hamiltonian Eqs.~(\ref{eq:dm-h}) and (\ref{eq:dm-hi}),
the wave function $|\Psi^V_E\rangle$ obtained by solving the Schr\"odinger equation in finite volume
must be of the following form
\begin{eqnarray}
|\Psi^V_E\rangle = |N^* \rangle \langle N^*|\Psi^V_E\rangle  + \sum_{i=1, n_c}\sum_{n_i=0,N-1}
|k_{n_i}\rangle \langle k_{n_i}|\Psi^V_E\rangle \, ,
\label{eq:psi-fv-comp}
\end{eqnarray}
where $|k_0\rangle, |k_1\rangle,\ldots, |k_{N-1}\rangle$ are the plane-wave states for a given
choice of $N$ momenta and $n_c$ is the number of meson-baryon channels considered.  Solving the
Schrödinger equation
\begin{eqnarray}
(H_0+H_I)|\Psi^V_E\rangle = E|\Psi^V_E\rangle \, ,
\label{eq:psi-fv}
\end{eqnarray}
in finite volume is then equivalent to finding the eigenvalues of the following matrix equation
\begin{eqnarray}
\det ([H_0]_{ N_c+1} + [H_I]_{N_c+1} - E[I]_{ N_c+1}) = 0 \, ,
\label{eq:det-1}
\end{eqnarray}
where $[I]_{N_c+1}$ is an $(N_c +1)\times(N_c+1)$ unit matrix with $N_c = N\times n_c$.

The matrix for the free Hamiltonian in Eq.~(\ref{eq:det-1}) takes the following form
\begin{eqnarray}
[H_0]_{N_c+1}&=&\left( \begin{array}{ccccccccc}
m_0                     & 0                            & 0
                        & \cdots                          & 0  & 0 &\cdots \\
0                       & \epsilon_1(k_0)      & 0
                        & \cdots                       & 0  & 0 & \cdots \\
0                       & 0                            & \epsilon_2(k_0)
                        & \cdots            &0                 & 0 & \cdots \\
0                       & 0                            & 0
                        & \ddots      & 0  & 0& \cdots \\
0                       & 0                            & 0
                        & \cdots                        & \epsilon_{n_c}(k_0) &0& \cdots \\
0                       & 0                            & 0
                        & \cdots   &0                       & \epsilon_1(k_1) & \cdots \\
\vdots                  & \vdots                       & \vdots
                        & \vdots                       & \vdots  &\vdots              & \ddots
\end{array} \right) \, , \nonumber
\end{eqnarray}
where $m_0$ is the mass of the bare $N^*$ state, and
\begin{eqnarray}
\epsilon_i(k_n) = E_{M_i}(k_n)+E_{B_i}(k_n)\, .
\end{eqnarray}
Here $E_{M_i}(k_n)$ and $E_{B_i}(k_n)$ are the free energies of the meson ($M$) and baryon ($B$) in
the $i$-th channel, respectively.  The $(N_c+1)\times(N_c+1)$ matrix for the interaction
Hamiltonian Eq.~(\ref{eq:dm-hi}) is
\begin{eqnarray}
[H_I]_{N_c+1}&=&\left( \begin{array}{ccccccccc}
0                       & g^V_{1}(k_0)               & g^{V}_{2}(k_0)
                        &\cdots& g^{V}_{n_c}(k_0)               & g^{V}_{1}(k_1)              & \cdots \\
g^{V}_{1}(k_0)   & v^V_{1,1}(k_0, k_0)   & v^{V}_{1,2}(k_0, k_0)
                        & \cdots & v^{V}_{1,n_c}(k_0, k_0)   & v^{V}_{1,1}(k_0, k_1)   & \cdots \\
g^{V}_{2}(k_0) & v^{V}_{2,1}(k_0, k_0) & v^{V}_{2,2}(k_0, k_0)
                        & \cdots& v^{V}_{2,n_c}(k_0, k_0) & v^V_{2,1}(k_0, k_1)   & \cdots \\
\vdots                  & \vdots         &\vdots
                        &\ddots          &\vdots &\vdots &\cdots \\
g^{V}_{n_c}(k_0)   & v^{V}_{n_c,1}(k_0, k_0)   & v^{V}_{n_c,2}(k_0, k_0)
                        & \cdots& v^{V}_{n_c,n_c}(k_0, k_0)   & v^{V}_{n_c,1}(k_0, k_1)   & \cdots \\
g^V_{1}(k_1) & v^{V}_{1,1}(k_1, k_0) & v^{V}_{1,2}(k_1, k_0)
                        &\cdots&  v^{V}_{1,n_c}(k_1, k_0) & v^{V}_{1,1}(k_1, k_1)   & \cdots \\
\vdots                  & \vdots                       & \vdots
                        & \vdots                       & \vdots         &\vdots         & \ddots
\end{array} \right)\, , \label{eq:3cmx}
\end{eqnarray}
with
\begin{eqnarray}
g^{V}_{i}(k_n)&=&\sqrt{\frac{C_3(n)}{4\pi}}\left(\frac{2\pi}{L}\right)^{3/2}
g_{N^*,i}(k_n)\, ,\\
v^{V}_{i,j}(k_{n_i},k_{n_j})&=&\sqrt{\frac{C_3(n_i)}{4\pi}}
\sqrt{\frac{C_3(n_j)}{4\pi}}\left(\frac{2\pi}{L}\right)^3 v_{i,j}(k_{n_i},k_{n_j}) \, ,
\label{eq:vfin-1}
\end{eqnarray}
where $C_3(n)$ is the number of degenerate states with the same magnitude $k_n=|\vec{k}_n|$.  By
solving Eq.~(\ref{eq:det-1}), we then obtain the spectrum $(E_1, E_2, \cdots)$ for each
partial-wave and the corresponding wave function of the eigenstate $|\Psi^V_{E_i}\rangle$.

In practice, we follow Refs.~\cite{ad-1,ad-2} in using the partial-wave matrix elements of
$g_{N^*,i}$ and $v_{i,j}$ to solve Eqs.~(\ref{eq:det-1}) through (\ref{eq:vfin-1}).  Thus the spin
information of particles is already included in the Hamiltonian matrix, which is of the same form
as for spinless particles.  We also neglect the contribution from higher partial waves which will
mix with S or P wave in defining the matrix equations. Accordingly, we only consider the pure P
wave contributions in the calculations of the spectra for the $P_{11}$ and $P_{33}$ channels.  With
these simplifications, only the Zeta function $Z_{00}(1,q^2)$ is needed to use the L\"uscher
formula to calculate the phase shifts from the predicted spectrum, as described below.  The
validity of this procedure has been established in Refs.~\cite{ad-1,ad-2}.

{}For the single channel $n_c=1$ case, the L\"uscher \cite{luch-1} formalism gives a phase shift
$\delta(E)$ for each energy $E$ of the predicted spectrum by
\begin{eqnarray}
\delta(E)&=&-\tan^{-1}\left(-\frac{q\pi^{3/2}}{Z_{00}(1;q^2)}\right)+ n\pi
\label{eq:luch1}
\end{eqnarray}
where $q=\frac{kL}{2\pi}$ is evaluated in terms of the three-momentum $k$ for the energy
$E=E_N(k)+E_\pi(k)$ of the spectrum, and $Z_{00}(1;q^2)$ is the generalized Zeta function.  The
formalism for two-channels was developed in Ref.~\cite{luch-2} and for the general multi-channel
case in Ref.~\cite{luch-3}.

With the eigenstate $|\Psi^V_{E_\alpha}\rangle$ (spin index omitted) in the rest frame of the
$N^*$, which is of the form of Eq.~(\ref{eq:psi-fv-comp}), {}from solving Eq.~(\ref{eq:det-1}), we
can calculate the probability of finding the bare state $N^*$:
\begin{eqnarray}
p^V_{N^*}(E_\alpha,L)= \left |\langle N^*|\Psi^V_{E_\alpha}\rangle \right |^2
\label{eq:prob-pin-fv}
\end{eqnarray}
As we will show explicitly in section III, $p^V_{N^*}(E_\alpha)$ is not a smooth function of
$E_\alpha$.  We therefore define the following energy-averaged form
\begin{eqnarray}
P_{N^*}^V(E^{\rm ave}_k, \Delta E, L) =\frac{1}{Z^V}\frac{1}{\Delta E}
\left [ \sum_{E^{\rm ave}_k-\frac{\Delta E}{2} \leq E_\alpha\leq E^{\rm ave}_k+\frac{\Delta E}{2}}
  p^V_{N^*}(E_\alpha,L) \right ] \, ,
\label{eq:pe-ave}
\end{eqnarray}
where
\begin{eqnarray}
Z^V=\sum_{\alpha} p_{N^*}^V(E_\alpha,L)\, ,
\label{eq:pe-norm}
\end{eqnarray}
which averages over states within a range $\Delta E$ centered at $E^{\rm ave}_k$.
From the above definitions, we have
\begin{eqnarray}
\sum_{k=1,N_E}P_{N^*}^V(E^{\rm ave}_k,L)\, \Delta E = 1 \, ,
\end{eqnarray}
where $N_E$ is the number of values, $(E^{\rm ave}_1, E^{\rm ave}_2,\cdots)$, chosen in the range
of the predicted spectrum used to obtain the energy-averaged values.
Obviously, $P_{N^*}^V(E^{\rm ave}_k,L)$, as defined in Eq.~(31), can have a well defined dependence on
$E^{\rm ave}_k$ only when there exists values of $E_i$ to cover the interval $\Delta E$ for each
chosen $E^{\rm ave}_k$.  From the spectrum calculated as a function of $L$, as will be shown in
Figs.~\ref{fg:sl} and \ref{fg:spect-p11}, it is straightforward to see that larger $L$ is required
in order to have a smooth $P_{N^*}^V(E^{\rm ave}_k,L)$, with a small $\Delta E$.  In the calculations to
be presented in the next two sections, we find that $L \times \Delta E \sim 4$ will yield a
well-defined function of $P_{N^*}^V(E^{\rm ave}_k,L)$.
With this relation in mind, we simplify our notation for $P_{N^*}^V(E, \Delta E, L)$
to $P_{N^*}^V(E, L)$.
Here we note that Eqs.~(\ref{eq:psi-fv}) and (\ref{eq:prob-pin-fv}) are the finite-volume versions
of Eqs.~(\ref{eq:psi-inf}) and Eq.~(\ref{eq:prob-i}) in infinite volume.  Thus it is reasonable to
assume that $P_{N^*}^V(E^{\rm ave}_k,L) $ can be compared with $P_{N^*}(E)$, defined by Eq.~(\ref{eq:pe-all}),
for infinite volume.  This will be demonstrated explicitly in the next section.

\section{One-channel dynamical model}
We first consider the dynamical Hamiltonian constructed in Ref.~\cite{sl}.  It has only one $\pi N$
channel and one bare $\Delta$ state in Eqs.~(\ref{eq:dm-h}) and (\ref{eq:dm-hi}).  By using
Eqs.~(\ref{eq:t-tot}) through (\ref{eq:selfe}) with $i=j=1=\pi N$, the $\pi N$ scattering amplitudes can be
calculated for each partial-wave.  The parameters of this model (the SL model) are determined by
fitting the data for the empirical $S$ and $P$ partial-wave amplitudes up to invariant mass $W=1.3$
GeV.  The fits to the data are shown in Fig.~\ref{fg:phase-sl}.

The potential $v_{\pi N, \pi N}$ for this single-channel dynamical model is based on the
meson-exchange mechanism.  This is essential to reduce the uncertainties in determining the partial
wave amplitudes from the data, which have unavoidable systematic and statistical errors. In
addition, the extracted $\Delta$ (1232) resonance parameters can be interpreted theoretically in
terms of a bare state surrounded by meson cloud.


\begin{figure}[t] \vspace{-0.cm}
\begin{center}
\includegraphics[width=0.9\columnwidth, clip]{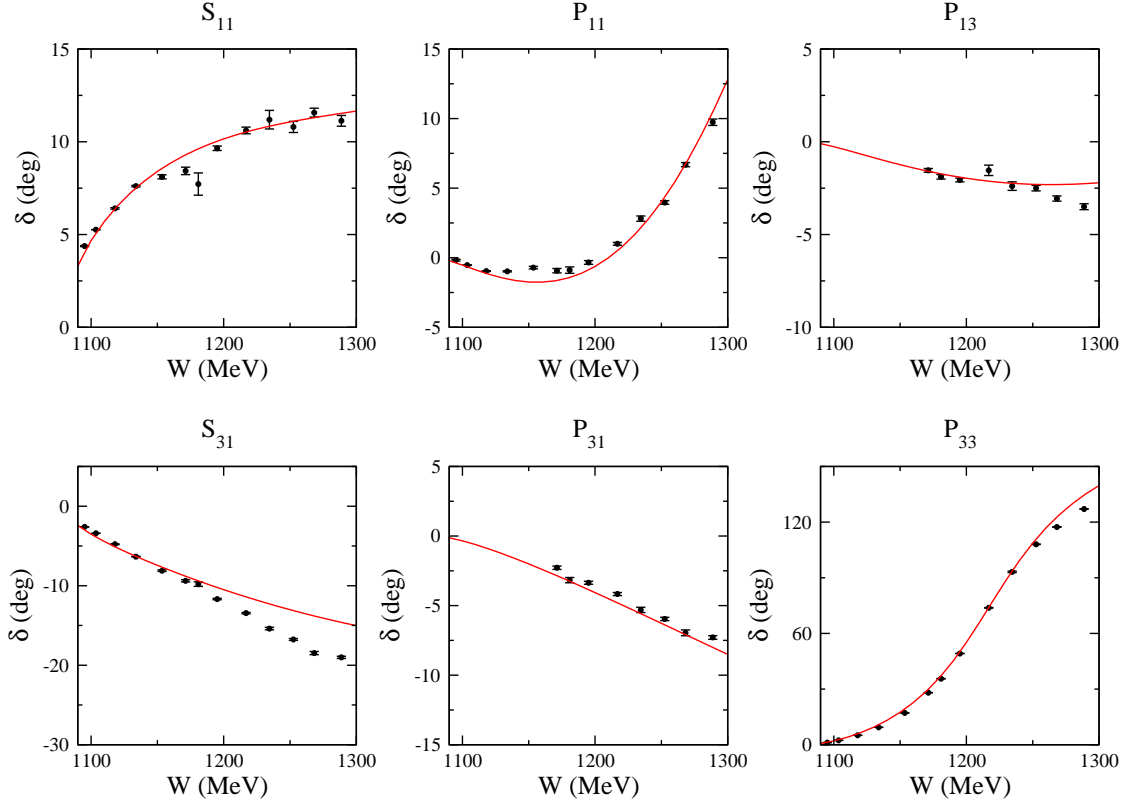}
\caption{ Phase shifts from the SL model \cite{sl} are compared with data in the $S$ and $P$
  partial-wave amplitudes.  Panels are labeled by $L_{2I\,2J}$.}
\label{fg:phase-sl}
\end{center}
\end{figure}

\begin{figure}[tbp] \vspace{-0.cm}
\begin{center}
\includegraphics[width=0.49\columnwidth]{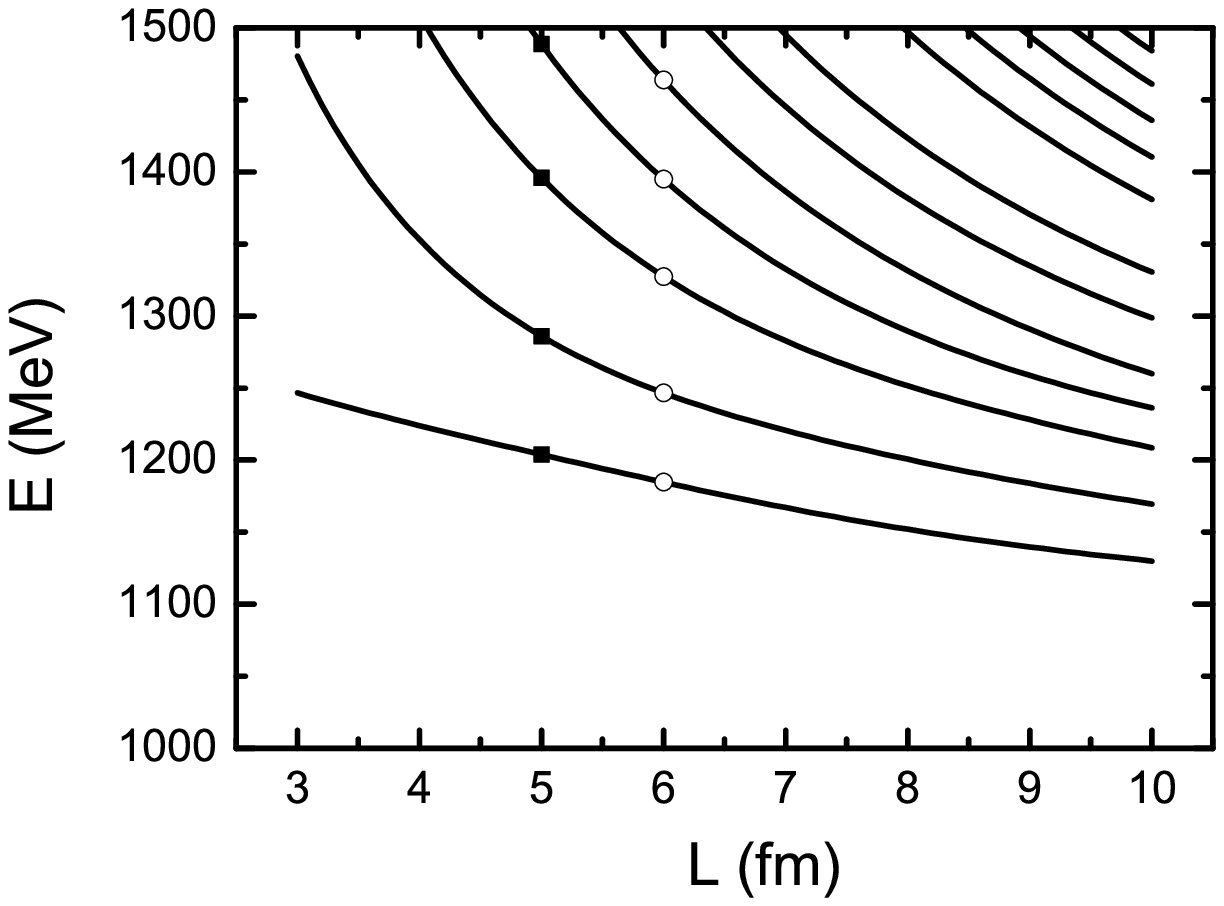}
\includegraphics[width=0.49\columnwidth]{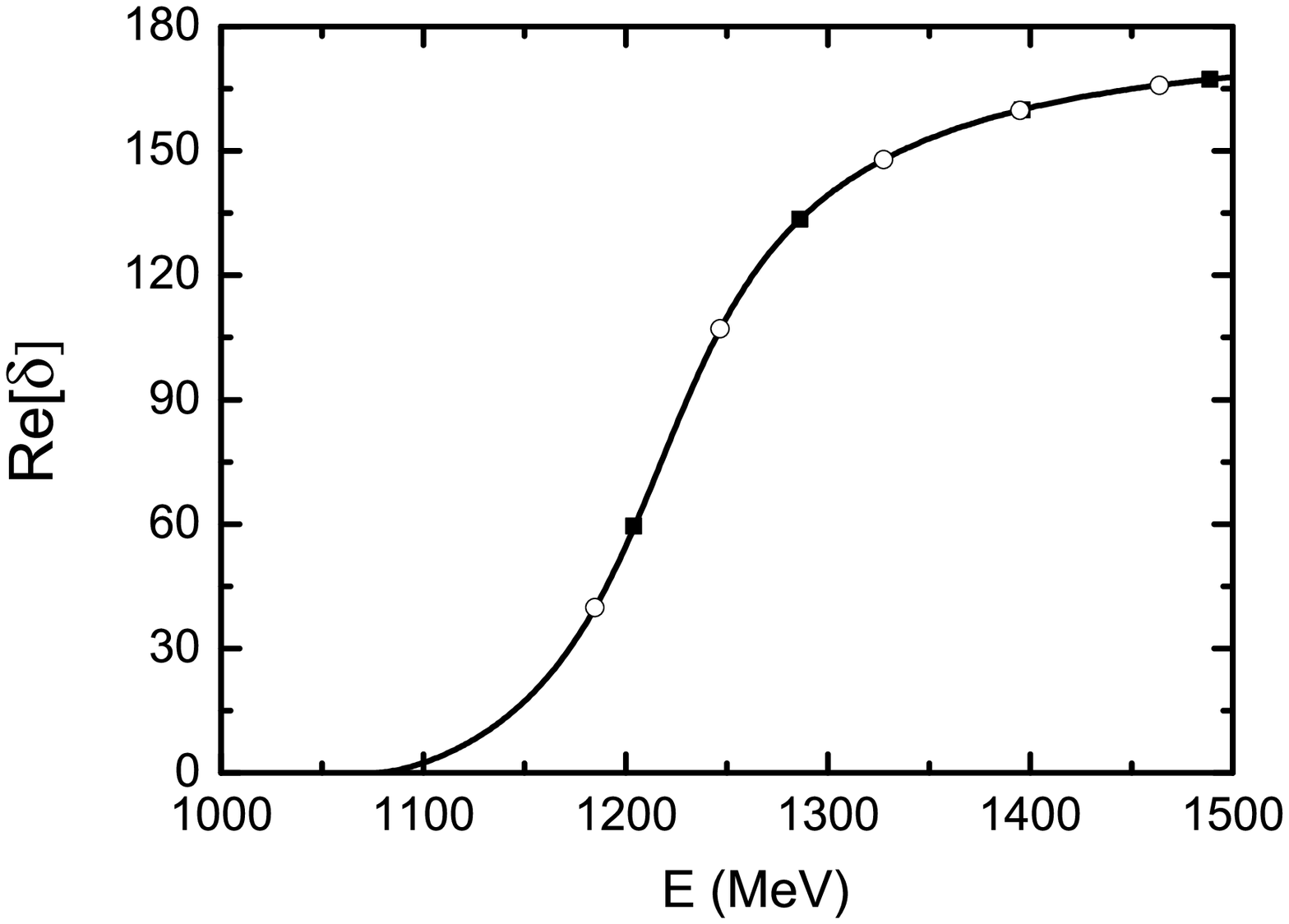}
\caption{(left) The finite-volume spectrum obtained via the FVH method through fits to the
  partial-wave scattering amplitudes is plotted as a function of the spatial lattice length $L$.
  The spectrum of states obtained at $L=5$ (full points) and 6 (open points) fm are used in the
  L\"uscher formalism to predict the experimental phase shifts in the right-hand panel.  (right)
  The phase shifts in the $P_{33}$ partial wave of the $\pi N$ system.  Full points and open points
  are obtained by applying the L\"uscher formalism to the finite volume spectra of the left-hand
  plot at $L=5$ and 6 fm respectively.  For reference the solid curve is that of the $P_{33}$
  channel displayed in Fig.~\ref{fg:phase-sl}, obtained through the fit of the SL model \cite{sl}
  to the partial-wave scattering amplitudes.  }
\label{fg:sl}
\end{center}
\end{figure}

Using the SL Hamiltonian as described above and solving Eqs.~(\ref{eq:det-1}) through
(\ref{eq:vfin-1}) with $n_c=1$, we obtain the finite volume spectrum in the $\pi N$ $P_{33}$
partial-wave.  This finite-volume spectrum is plotted in the left-hand panel of Fig.~\ref{fg:sl} as
function of the spatial lattice length $L$.

With this knowledge of the finite-volume spectrum of states having the quantum numbers of the
$\Delta$, one can then use the L\"uscher relation of Eq.~(\ref{eq:luch1}) to predict the $P_{33}$ phase shift for
each of the energy levels of the predicted spectrum.   These results are reported in the right-hand panel
of Fig.~\ref{fg:sl} as open and full points.  Open points have their origin in the finite-volume
spectrum obtained at $L=6$ fm while the full points follow from the spectrum at $L=5$ fm as
indicated in the left-hand panel of Fig.~\ref{fg:sl}.

For reference, the solid curve illustrated in the right-hand panel of Fig.~\ref{fg:sl} is that of
the $P_{33}$ channel displayed in Fig.~\ref{fg:phase-sl}, obtained through the fit of the SL model
to the partial-wave scattering amplitudes.

We note that the phase shifts calculated from each point of the spectra at $L=5$ and 6 fm agree
with the solid curve which is consistent with the experimental values and tied to the finite-volume
spectrum via the FVH model.  Thus the FVH method is equivalent to the use of L\"uscher's formula in
relating the finite-volume spectrum to the scattering phase shifts determined by the experimental
data.  This is in agreement with the findings of Refs.~\cite{ad-1,ad-2}, which used separable
potentials.

The spectrum shown in the left side of Fig.~\ref{fg:sl} can be used to examine whether
the experimental data, brought to the finite volume of the lattice via the SL model, is consistent
with LQCD, and vice-versa.
However, this comparison does not necessarily test the physics considered in the formulation of the
model.
As demonstrated in Ref.~\cite{lwk}, when the experimental data are complete and of very high
accuracy, the predicted $\Delta$ resonance properties are independent of the model when the
model(s) considered describe the data accurately.
As illustrated in Fig.~\ref{fg:phase-sl}, the SL model has met this condition reasonably well.

We now turn to examine the probability, $P_{N^*}(E)$ of Eq.~(\ref{eq:pe-all}) for the $\Delta$
resonance.  $P_\Delta(E)$ describes the probability to find the bare $\Delta$ in the $\pi N$
scattering wave function.  Within the SL model, the predicted $P_\Delta(E)$ and the resonant cross
section, $\sigma^{\rm res}(E)$, are compared in Fig.~\ref{fg:pe-crs-p33}.
We see that they have the same resonant structure near $E=1232$ MeV.  This is not surprising, as
can be seen by comparing the expressions of Eqs.~(\ref{eq:sigma-r}) and (\ref{eq:prob-pin-ifv}).
The results shown in Fig.~\ref{fg:pe-crs-p33} indicate that the predicted $P_\Delta(E)$ contains
the information of the extracted $\Delta$ resonance projected onto the physical real-$E$ axis.

We next use Eq.~(\ref{eq:prob-pin-fv}) to calculate $p^V_{\Delta}(E,L)$ which is the probability of
finding the bare $\Delta$ in the eigenstate $|\Psi^V_E>$ of the Hamiltonian in finite volume.  We
see in Fig.~\ref{fg:SLwave2} that the calculated $p^V_{\Delta}(E,L)$ is not a smooth function of
$E$ for each $L$.  As demonstrated in Appendix B within an exactly soluble model, the fluctuations
are a mathematical consequence of the quantization condition in finite volume.  Nevertheless, the
general structure of $p^V_{\Delta}(E, L)$ has a resonant shape as $L$ increases.  We then find that
the energy-averaged $P_\Delta^V(E,  L)$, as defined by Eq.~(\ref{eq:pe-ave}), is more
useful as a comparison with $P_\Delta(E)$ from infinite volume.  This can be seen in
Fig.~\ref{fg:SLPE}, where $P^V_{\Delta}(E,  L)$ clearly approaches $P_\Delta(E)$ as the
lattice size, $L$, increases.

Our results suggest that it will be interesting to calculate the analogue of $P_{N^*}^V(E,  L)$
directly from LQCD.
The formalism developed herein establishes a bridge between $P_\Delta(E)$ of the SL model in the
infinite volume of experiment and the finite-volume analogue.
It will be fascinating to explore the possibility of a similar quantity evaluated directly in terms
of the underlying dynamics of QCD.

Obtaining a $P_\Delta^V(E,L)$ in LQCD for large $L$ is very difficult.  Nevertheless, the results
shown in Fig.~\ref{fg:SLPE} suggest that $P_\Delta^V(E,L)$ can qualitatively reproduce the shape of
$P_\Delta(E)$ even for $L=3$ fm.  We will discuss possible calculations of $P_\Delta^V(E,L)$ in
section V.

\begin{figure}[tbp] \vspace{-0.cm}
\begin{center}
\includegraphics[width=0.7\columnwidth]{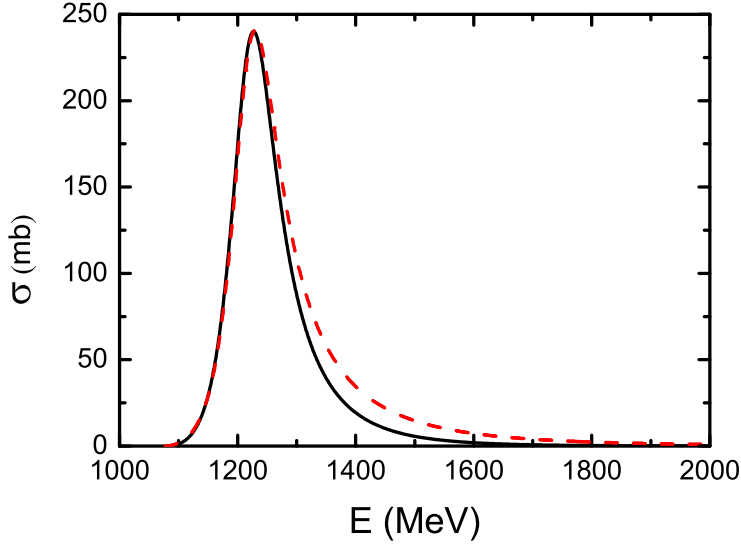}
\caption{Comparison of the energy-dependence of the resonant cross section of $\pi N$ elastic
  scattering in the $P_{33}$ partial-wave channel, $\sigma^{\rm res}(E)$, (black solid curve) and
  the probability to find the bare $\Delta$ in the $\pi N$ scattering wave function, $P_\Delta(E)$
  (red dashed curve), normalized at the peak.}
\label{fg:pe-crs-p33}
\end{center}
\end{figure}

\begin{figure}[tbp] \vspace{-0.cm}
\begin{center}
\includegraphics[width=0.8\columnwidth]{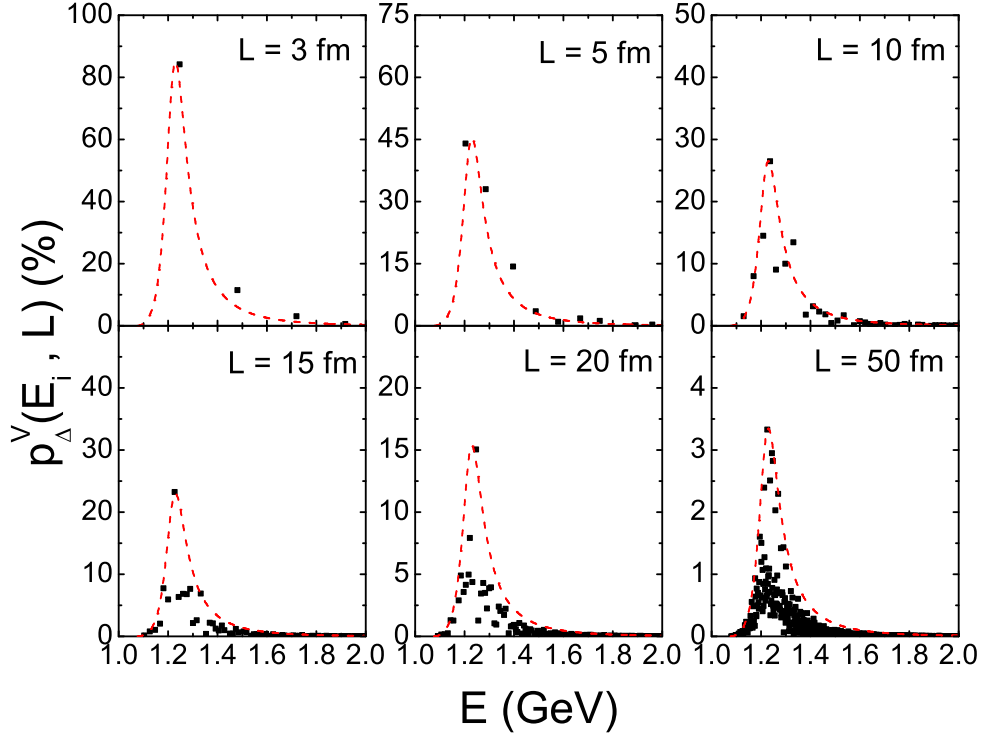}
\caption{The overlap probability, $p_\Delta^V(E_i,L) = |\langle \Delta | \psi^V(E_i) \rangle |^2$, of
  the bare $\Delta$ with the finite-volume energy eigenstate for SL model is shown as solid square points , at $L=3, 5, 10, 15, 20, 50$ fm.
  The red dashed curves show the infinite-volume $P_\Delta(E)$ normalized at the peak.
  }
\label{fg:SLwave2}
\end{center}
\end{figure}

\begin{figure}[tbp] \vspace{-0.cm}
\begin{center}
\includegraphics[width=0.8\columnwidth]{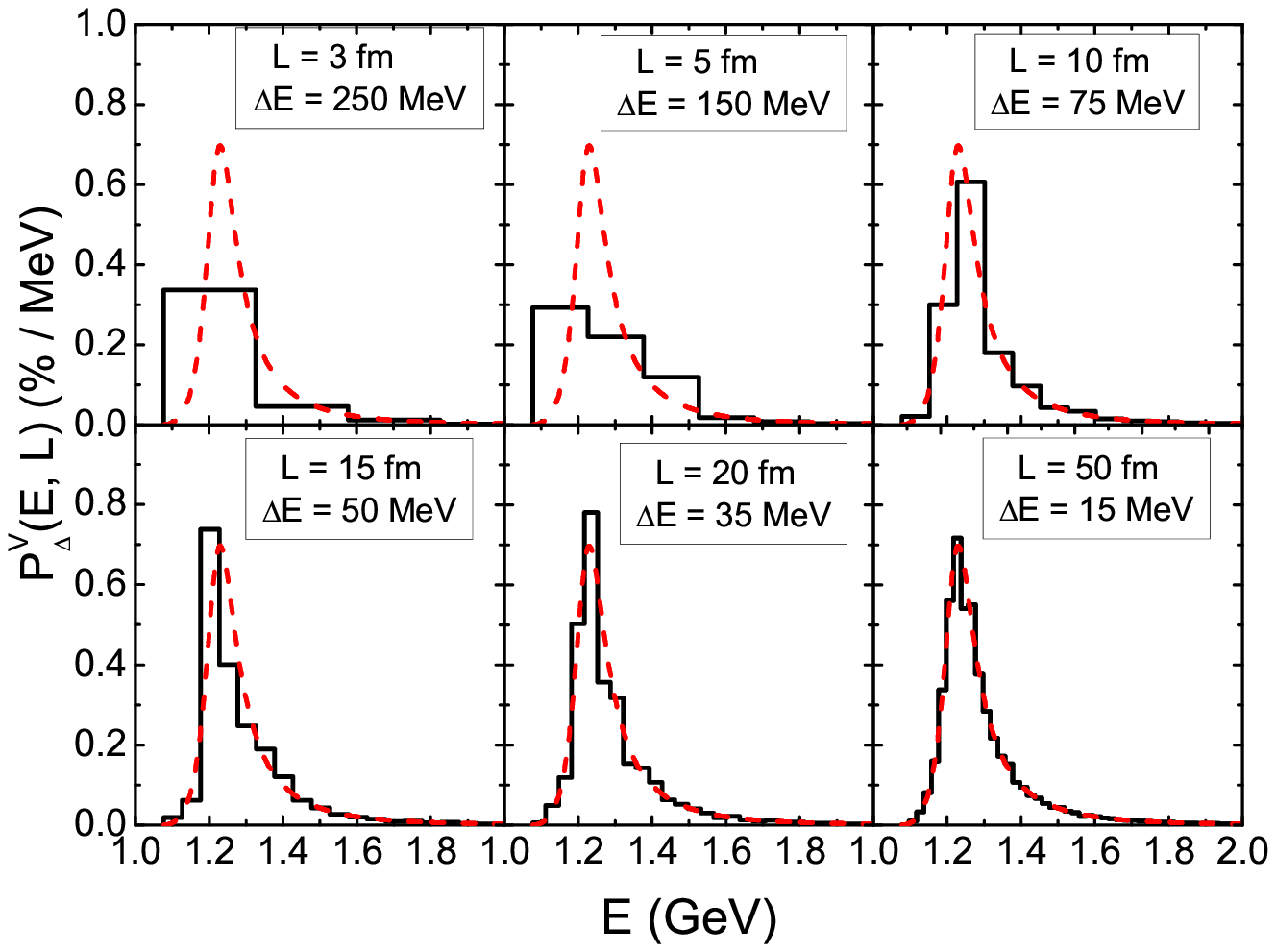}
\caption{The probability $P_\Delta^V(E, L)$ (black solid line) and $P_\Delta(E)$ (red
  dashed curve) of the FVH SL model at $L=3, 5, 10, 15, 20,$ and 50 fm.}
\label{fg:SLPE}
\end{center}
\end{figure}

\section{Three-channel dynamical model}
In this section, we consider a three-channel model in the form of Eqs.~(\ref{eq:dm-h}) and
(\ref{eq:dm-hi}).  
It includes the $\pi N$, $\pi \Delta$, and $\sigma N$ channels, where
$\Delta$ and $\sigma$ in the latter two channels are both treated as
stable particles. 
The meson-exchange two-body interactions $v_{i,j}$ with $i,j=\pi N, \pi \Delta, \sigma N$ are taken from the ANL-Osaka
Hamiltonian \cite{anl-osaka}, and one bare state is included in each partial wave except $S_{11}$ and $P_{31}$.
Their parameters are adjusted, along with the vertices $g_{N^*,i}$ with $i=\pi N, \pi \Delta, \sigma N$, to
fit the $S$- and $P$- partial-wave $\pi N$ empirical amplitudes \cite{said} up to invariant mass
$W=1.6$ GeV.  We see in Fig.~\ref{fg:amp-sp} that the fits are reasonable.  The only exception is
the $S_{11}$ partial wave, which is known to have a large coupling with the $\eta N$ channel and
therefore cannot be fitted well in this model.  Herein, we focus on the results in the $P_{11}$ and
$P_{33}$ partial waves.

\begin{figure}[htbp] \vspace{0.2cm}
\begin{center}
\includegraphics[width=0.8\columnwidth, clip]{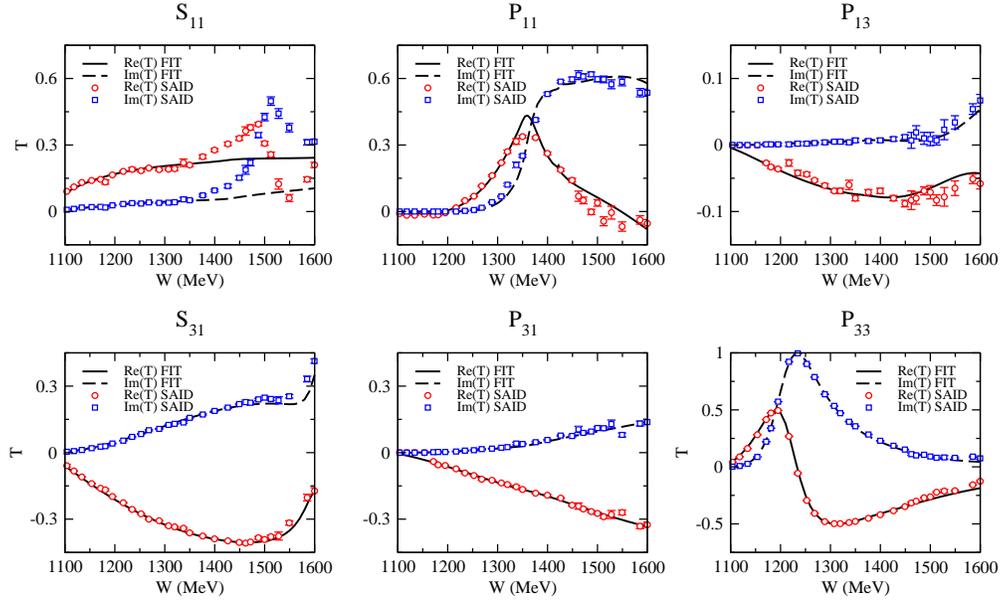}
\caption{Fits to the empirical data \cite{said} for the $\pi N$ partial-wave amplitudes.  Panels
  are labeled by $L_{2I\,2J}$.  With the exception of the $S_{11}$ partial wave, where the $\eta N$
  channel is required, the model describes the partial-wave amplitudes well.}
\label{fg:amp-sp}
\end{center}
\end{figure}

By solving Eq.~(\ref{eq:det-1}) in finite volume, we obtain the spectrum for each partial-wave. The
results for the $P_{33}$ and $P_{11}$ partial waves are shown in Fig.~\ref{fg:spect-p11}. It is
interesting to note that the predicted spectrum for $P_{33}$ partial wave (left-hand panel) from
the three-channel model agrees well with the solid squares taken from the spectrum of the
single-channel SL model reported in Fig.~\ref{fg:sl}.  This indicates that the predicted
finite-volume spectra are not sensitive to the details of the Hamiltonian provided the models agree
on the predicted scattering amplitudes.  This is in agreement with the findings in a study of
two-channel cases in Ref.~\cite{ad-2}.  The calculated spectra for the $P_{11}$ partial wave are
shown as the solid curves in the right-hand panel of Fig.~\ref{fg:spect-p11}.

\begin{figure}
\begin{center}
\includegraphics[width=0.49\columnwidth]{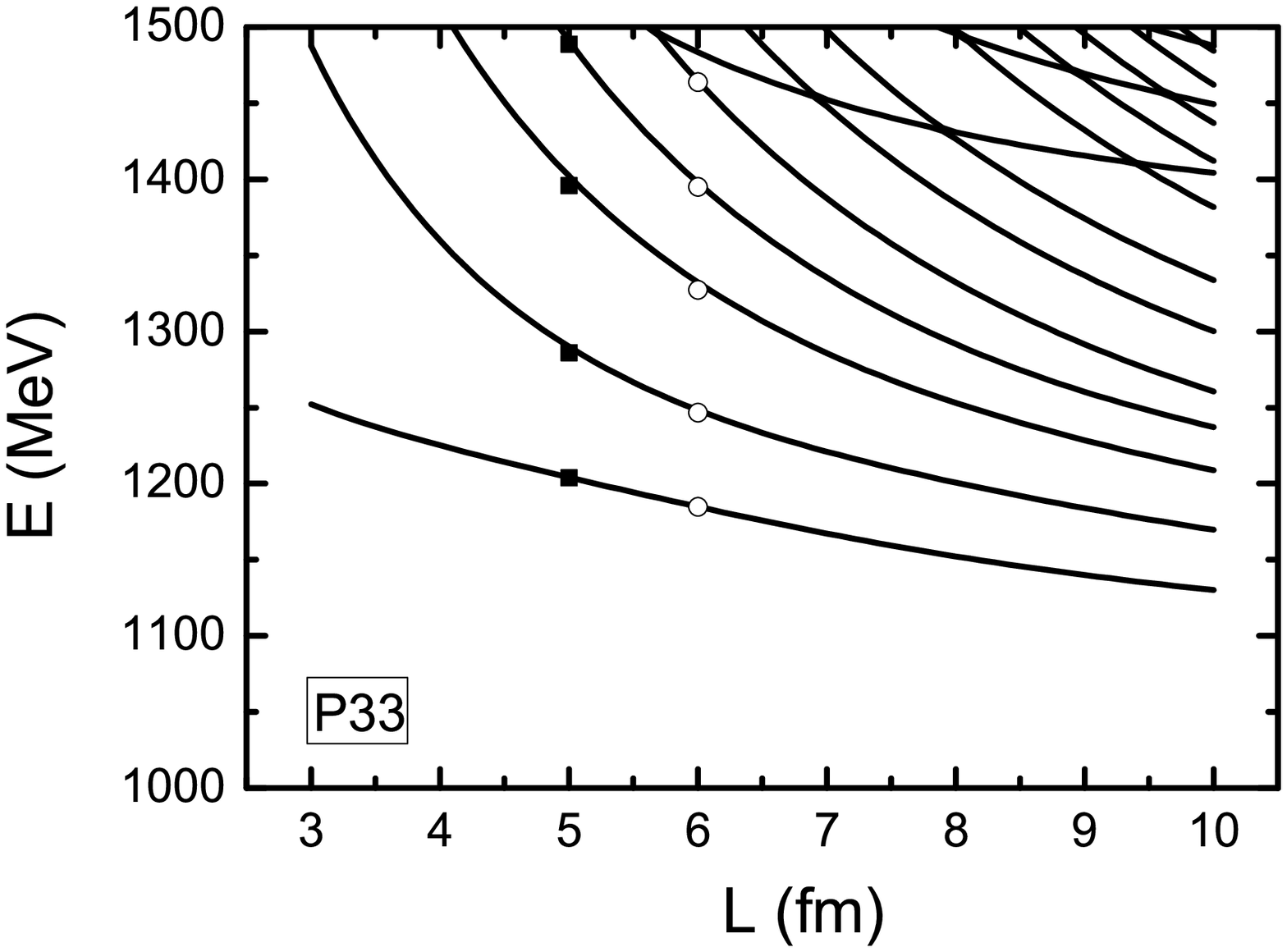}
\includegraphics[width=0.49\columnwidth]{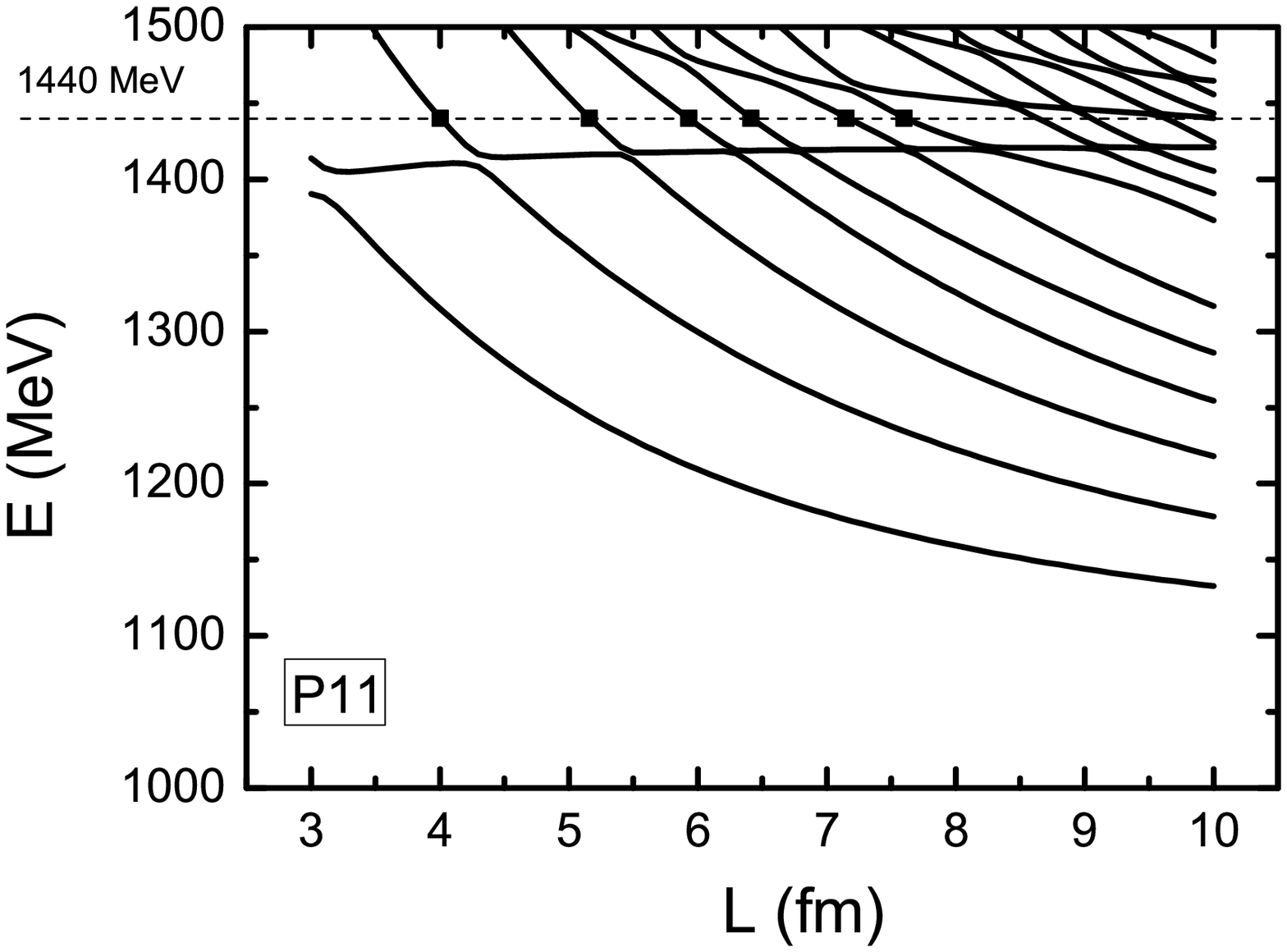}
\caption{The finite-volume spectrum for the $P_{33}$ (left) and $P_{11}$ (right) partial waves,
  calculated from the three-channel model incorporating $\pi N$, $\pi\Delta$ and $\sigma N$,
  are plotted as a function of the spatial lattice length $L$.
We note the finite line width hides some of the weakly coupled avoided level crossings,
particularly in the left-hand panel for the $P_{33}$ partial wave.  Avoided level crossings of
$\sigma N$, $\pi N$ and $\pi\Delta$ channels are readily apparent in the right-hand panel for the
$P_{11}$ partial wave.
The squares in the left-hand $P_{33}$ panel are taken from Fig.~\ref{fg:sl} for the single-channel
model and illustrate the independence of the finite-volume spectrum from the model, when both
models describe the empirical partial-wave scattering data well.
In the right-hand panel, the six solid squares on the dashed line indicate the six lattice volumes
that need to be considered in order to constrain the relations of the multi-channel L\"uscher
formula.  }
\label{fg:spect-p11}
\end{center}
\end{figure}

The L\"uscher formalism has been extended in Ref.~\cite{luch-3} to the general multi-channel
system.  By choosing the normalization to relate the T-matrix elements to S-matrix elements by
$S_{\alpha,\beta}(E)=\delta_{\alpha,\beta} - 2i T_{\alpha,\beta}(E)$, the formula given in
Ref.~\cite{luch-3} for the constructed 3-channel model can be written explicitly as :
\begin{eqnarray}
\det[M(E,L)]=0
\label{eq:detf}
\end{eqnarray}
where
\begin{eqnarray}
&&M(E,L)\nonumber\\
&=&\left(
\begin{array}{ccc}
T_{\pi N,\pi N}(E)+C_{\pi N,\pi N}(L,E)     & T_{\pi N,\pi\Delta}(E)   & T_{\pi N,\sigma N}(E) \\
T_{\pi \Delta, \pi N}(E)  & T_{\pi \Delta,\pi\Delta}(E)+C_{\pi\Delta,\pi\Delta}(L,E) &T_{\pi\Delta,\sigma N}(E) \\
T_{\sigma N, \pi N}(E)    & T_{\sigma N,\pi \Delta}(E)& T_{\sigma N,\sigma N}(E)+C_{\sigma N,\sigma N}(L,E)
\end{array}
\right), \nonumber
\end{eqnarray}
and
\begin{eqnarray}
 C_{\alpha,\alpha}(L,E)&=&\frac{iq_\alpha(L)}{q_\alpha(L)-4\sqrt{\pi}Z_{00}(1;q_\alpha(L))},
\label{eq:detf-1}
\end{eqnarray}
and $q_\alpha(L)=k_\alpha L/(2\pi)$ is defined by the on-shell momentum $k_\alpha$ of total energy $E$ in channel
$\alpha$.
Because of symmetries and the unitary conditions, only six of the total 12 real numbers needed to
specify all six of the complex $T_{\alpha,\beta}(E)$ matrix elements are independent.  Thus we need
to get six relations from Eqs.~(\ref{eq:detf}) through (\ref{eq:detf-1}) at each $E$ to relate the
spectrum to the scattering amplitudes shown in Fig.~\ref{fg:amp-sp}.  In the rest frame, this means
that we need to perform LQCD calculations at six different values of $L$. For $E=1440$ MeV, this is
indicated by the six solid squares on the dashed line at the intersections of the solid curves in the
left-hand panel of Fig.~\ref{fg:spect-p11}.  Clearly, this constitutes an extremely difficult and
time consuming LQCD calculation.

In addition, because the Roper $N^*(1440)$ is broad, one needs to get LQCD data over a range of
order 500 MeV around 1440 MeV to construct a model, such as the K-matrix model employed in
Refs.~\cite{jlab-lqcd-1,jlab-lqcd-2}, in order to extract the resonance parameters by analytic
continuation to the complex energy plane.

On the other hand, the information on the Roper $N^*(1440)$ resonance has been encoded in the
three-channel Hamiltonian by fitting the empirical $\pi N$ scattering amplitudes \cite{said}, as
shown in Fig.~\ref{fg:amp-sp}.  Therefore the spectrum from the finite-volume Hamiltonian method at
any given $L$ is sufficient to understand and test LQCD results.  This is a significant
advantage of the finite-volume Hamiltonian method over using the L\"uscher formalism in resolving
the dynamics of LQCD calculations through the investigation of nucleon resonances.

\begin{table}[b]
\caption{\label{tab:pole} The $P_{33}$ and $P_{11}$ resonance pole masses ($M_R$) extracted from
  the three-channel model.  Each resonance pole mass is listed as
  $\bm{(}\textrm{Re}(M_R),-\textrm{Im}(M_R)\bm{)}$.
  Experimental values are from Ref.~\cite{pdg}.
  The masses for the input bare $N^*$ states are also listed in the third column.  }
\begin{ruledtabular}
\begin{tabular}{ccccc}
$L_{2I\,2J}$  & Resonance     & Pole Masses (MeV)& Experiment (MeV)  & Bare Masses (MeV) \\
\hline
$P_{33}$      &$\Delta(1232)$ & (1212, 53)       & (1209-1211, 49-51)      &     1470            \\
\noalign{\smallskip}
$P_{11}$      &$N^*(1440)$    & (1354, 38)       & (1350-1380, 80-110)     &     2100   \\
              &$N^*(1710)$    & (1717, 73)       & (1670-1770, 40-190)     &                     \\
\end{tabular}
\end{ruledtabular}
\end{table}

We now investigate the resonances extracted within this three-channel model.  The extracted pole
positions and bare masses are listed in Table~\ref{tab:pole}.  The value of the resonance pole in
the $P_{33}$ channel is close to the value $M_R=1216.4\,\,-i\, 50.0$ MeV found in the SL model
\cite{sl}.  This is in agreement with the finding of Ref.~\cite{lwk} that the resonance extraction
is independent of the model, so long as the data near the resonance positions are very accurate and
fitted $precisely$. This is also evident in a comparison of the $P_{33}$ results in
Figs.~\ref{fg:sl} and \ref{fg:amp-sp} in the region 1100 MeV $\leq W \leq $ 1250 MeV.

Turning to the $P_{11}$ channel we have two poles with masses $M_{R_1} =1354.0\,\, - i\,38.0$ MeV
and $M_{R_2}=1717.0\,\,-i\,73.0$ MeV.
The situation is much more complicated in this case than for the $P_{33}$.  However, we find that $P_{N^*}(E)$
of Eq.~(\ref{eq:pe-tot}), which measures the probability of finding the bare state in the
meson-baryon scattering wave functions, still contains the information concerning the extracted
resonances.  This can be seen in Fig.~\ref{fg:pe-crs-p11}.  We find that $P(E)$ has a similar
energy-dependence to that of the resonant part of the elastic cross section, $\sigma^{\rm
  res}(E)$. In particular, the structure near $W=1400$ MeV, reflecting the broad Roper resonance on
the real-axis, is also seen in $P_{N^*}(E)$.

By using the wave function, $|\Psi^V_{E}\rangle$, obtained by solving Eq.~(\ref{eq:det-1}) for the
three-channel Hamiltonian in finite volume, we can calculate $P_{N^*}^V(E,L)$ using Eq.~(\ref{eq:pe-ave}).
We see in Fig.~\ref{fg:1b3cPE} that the energy-averaged $P_{N^*}^V(E,L)$ agrees very well with $P_{N^*}(E)$.
Thus $P_{N^*}(E)$ can also be used to check whether the extracted resonances are consistent with the
underlying QCD dynamics, provided $P_{N^*}^V(E,L)$ can be calculated for sufficiently large $L$.

\begin{figure}[tbp] \vspace{-0.0cm}
\begin{center}
\includegraphics[width=0.6\columnwidth, clip]{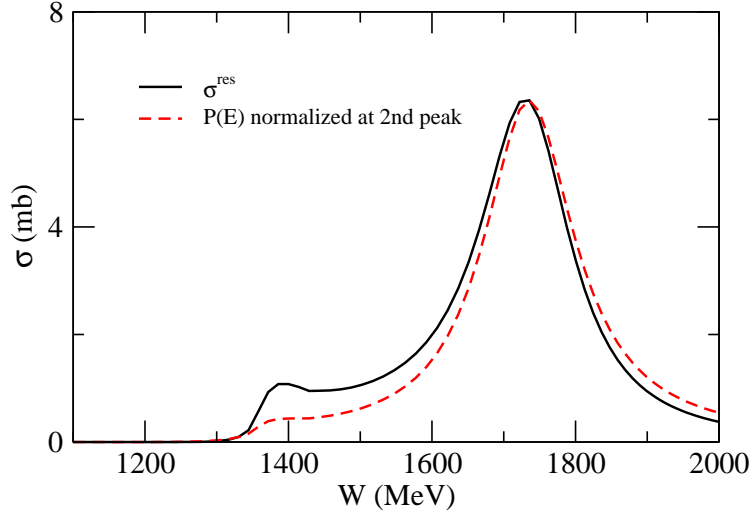}
\caption{Comparison of the energy-dependence of $\sigma_{res}$ and $P_{N^*}(E)$ for the $P_{11}$
  partial-wave channel, normalized at the peak.}
\label{fg:pe-crs-p11}
\end{center}
\end{figure}

\begin{figure}[tbp] \vspace{-0.cm}
\begin{center}
\includegraphics[width=0.8\columnwidth]{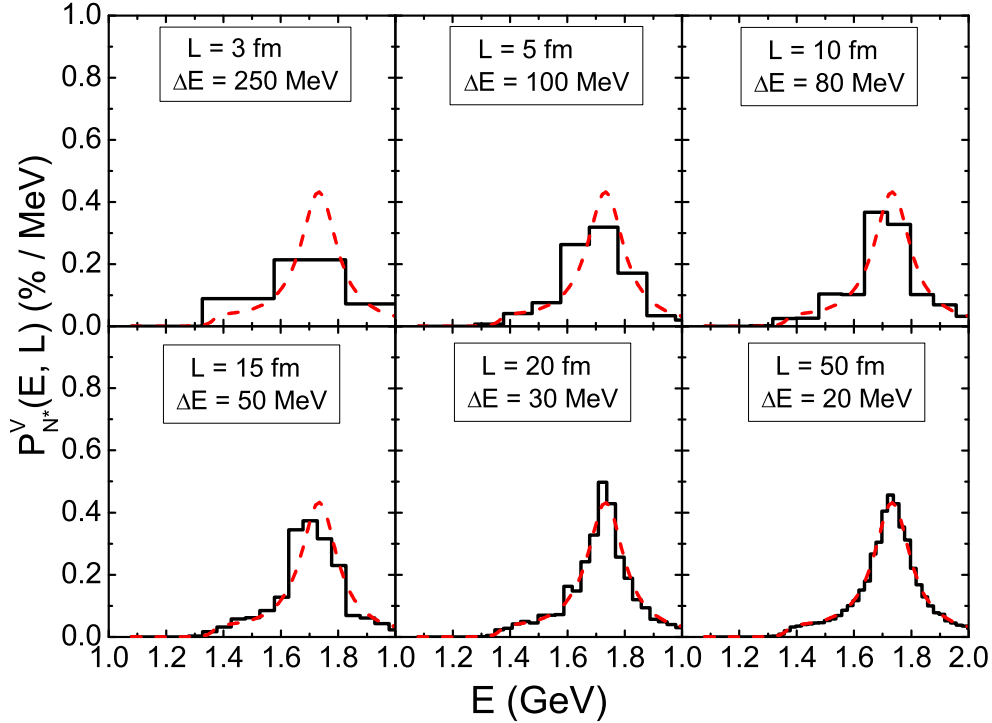}
\caption{The finite-volume $P^{V}_{N^*}(E, L)$ (black solid) and infinite-volume $P_{N^*}(E)$
  (red dashed) bare-state probabilities for the three-channel model at
  $L=3,\ 5,\ 10,\ 15,\ 20$, and 50 fm.}
\label{fg:1b3cPE}
\end{center}
\end{figure}

\section{LQCD calculations of $P^V_{N^*}(E, L)$}

Here we explore LQCD calculations of $P_{N^*}^V(E,L)$ and the extent to which measures can be
related to the bare states of the dynamical model.  Since $P_{N^*}^V(E,L)$ reflects the properties
of the resonance, the direct measurement of $P_{N^*}^V(E,L)$ from LQCD will provide the insight
needed for understanding the essence of resonance structure.  It holds the promise to further
elucidate the effective mechanisms of QCD dynamics and extend our knowledge of QCD.

There are fundamental QCD dynamics that support the concept of a hadronic quark core dressed by a
meson cloud.  A particularly illustrative example is that of coherent center domains in the vacuum
of QCD \cite{Stokes:2013oaa}.  Within the domains governed by the trace of the Polyakov loop,
color-singlet quark-antiquark pairs or three-quark triplets have a finite energy and are spatially
correlated.  These fundamental domains are thought to govern the size of the quark cores of hadrons
\cite{Stokes:2013oaa}.

Of course, there is some model dependence in the separation of an energy eigenstate into its core
or bare-state contribution and its associated meson-cloud contribution.
For example, in effective field theory this separation is governed by the scale of the
regulator \cite{Young:2002ib,Leinweber:2003dg} and in the power-counting regime of chiral
perturbation theory, the physics of the expansion is independent of the regulator
\cite{Hall:2010ai}.  The physics can be shifted from the core to the cloud through a change in the
regulator parameter value with no change in the renormalized low-energy coefficients.
However, when working beyond the power-counting regime, an intrinsic scale reveals itself through a
convergence in the values of the renormalized low-energy coefficients of the expansion
 \cite{Hall:2010ai,Hall:2012pk,Hall:2012yx,Hall:2013oga}.  For dipole regulators, a scale of $\sim
1$ GeV is found.  This intrinsic scale is associated with the finite size of the source of the
meson cloud and phenomenology suggests a scale of 0.8 GeV
\cite{Young:2001nc,Young:2002cj,Young:2002ib,Leinweber:2003dg,Leinweber:2004tc,Leinweber:2005bz,%
Leinweber:2006ug,Wang:2007iw,Wang:1900ta,Wang:2012hj,Wang:2013cfp}.

With this insight, one can attribute some physics to the baryon core and the balance to
the meson cloud.  This approach has been very successful in correcting the meson cloud of quenched
QCD to make precise full QCD predictions
\cite{Young:2001nc,Young:2002cj,Leinweber:2004tc,Leinweber:2005bz,Leinweber:2006ug,Wang:2007iw,%
  Wang:1900ta,Wang:2012hj,Wang:2013cfp}.  In this case the baryon core is held invariant between
quenched and full QCD and the artifacts of the quenched meson cloud are removed and replaced with
the full QCD cloud contribution.

In previous coupled-channel effective field theory studies of the $\Delta$(1232) resonance it has
been concluded that the $\Delta$(1232) resonance can be interpreted as a system made of a quark
core and a meson cloud.  Furthermore, the contributions from the quark core to the electromagnetic
$\gamma^* N \rightarrow \Delta$ form factors are found to be similar to the predictions from the
three-quark configurations within either the constituent three-quark model or models based on the
Dyson-Schwinger equations (DSE).  Since the meson cloud effects within the SL model are defined by
well-studied meson-exchange mechanisms and are strongly constrained by fitting the $\pi N$
scattering phase shifts in all partial-waves, this separation of the core and meson cloud is not
completely arbitrary.
In summary, there is ample evidence that the essential underlying mechanism of baryon structure is
that of a quark core surrounded by a meson cloud.

The results shown in Figs.~\ref{fg:SLPE} and \ref{fg:1b3cPE} establish a relationship between the
probabilities of finding the bare state in infinite volume, $P_{N^*}(E)$, and in finite volume,
$P_{N^*}^V(E,L)$.  The relationship enables a new exploration of connecting $P_{N^*}^V(E,L)$,
containing resonance information extracted from the $\pi N$ reaction data within a dynamical model,
to that obtained directly from lattice QCD.

Our hypothesis is that the probability of finding the bare state in a finite-volume eigenstate of
lattice QCD, $P_{N^*}^V(E,L)$, is related to the overlap of an appropriately smeared three-quark
lattice interpolating field with the lattice QCD eigenstates.  As there is some freedom in defining
this three-quark operator, it will be important to examine the parameter space as one selects an
operator that models the three-quark core.

For example, the spin-flavour nature of the interpolating field must be selected.  For
local three-quark operators, the choice for $N$ and $\Delta$ baryons is straight forward.  The
spin-flavour structure for the $\Delta$ is unique \cite{Chung:1981cc} and there is only one
spin-1/2 nucleon operator that overlaps significantly with the nucleon and its radial excitations
\cite{Leinweber:1994nm,Mahbub:2010rm,Mahbub:2013ala}.

Similarly, the source of the quark propagator is smeared out to provide a finite size for the
distribution of quarks within the quark core.  As detailed in Appendix A, the smearing is performed in an
iterative manner that gives rise to a Gaussian-shaped distribution with the size governed by the
number of iterations.  Radially excited cores can be constructed from a superposition of Gaussian
smeared sources to create a node
\cite{Mahbub:2010rm,Mahbub:2013ala,Roberts:2013ipa,Roberts:2013oea}.
%

It will be interesting to examine the dependence of $P_{N^*}^V(E,L)$ on this smearing extent.  In
selecting a range of interesting values one can consider the size of the hadron as measured in form
factors and draw on insight from the typical size of coherent center domains in the QCD vacuum.
It's well known that smaller smearing extents have better overlap with higher excited states of the
spectrum \cite{Mahbub:2010rm} and thus there is a relationship between the smearing extent and the mass
of the bare state.

We note that the discrete nature of the finite-volume LQCD spectrum prevents a determination
of $p_{N^*}^V(E,L)$ for arbitrary $E$.
LQCD can only calculate $P_{N^*}^V(E_\alpha,L)$, as defined in Eq.~(\ref{eq:prob-pin-fv}), for the
$\alpha$'th eigenstate, $|\Psi^V_{E_\alpha}\,\rangle$
\begin{eqnarray}
p^V_{N_{C}^*}(E_\alpha,L)= \left |\langle\, N_{C}^* \,|\, \Psi^V_{E_\alpha}\, \rangle \right |^2
\equiv \left |\lambda_C^{\alpha}\right |^2 \, .
\label{eq:lqcd-pv}
\end{eqnarray}
The task then is to define a bare or core state on the lattice $|\, N_{C}^* \, \rangle$.  To do
this we resort to the aforementioned local three-quark interpolating field, $\chi_C$, acting on the
QCD vacuum $|\Omega\rangle$.  In the rest frame of the state
\begin{eqnarray}
\lambda^{\alpha}_C \, u^\alpha(\vec 0) &=& \langle\, N_{C}^* \, \vert \, \Psi^V_{E_\alpha}\, \rangle
=\langle\, \Omega\, \vert\, \chi_C\,  \vert\, \, \Psi^V_{E_\alpha} \rangle\, .
\label{eq:lqcd-pv-1}
\end{eqnarray}
where $u^\alpha(\vec 0)$ is the zero-momentum Dirac spinor for state $\alpha$.  Here $N_{C}^*$ and
thus $\chi_C$ encode the spin, isospin and parity of the Core state $C$ under consideration.  This
can be the bare Nucleon, bare Roper, bare $N^*(1535)$, bare $\Delta$ and so on.
$|\Psi^V_{E_\alpha} \rangle$ is the $\alpha$'th lattice QCD eigenstate in the finite volume.  As an
example, consider the $[JTP]=[3/2,3/2,+]$, $\Delta^{++}$ state where there is only one local
three-quark operator transforming as a Rarita-Schwinger spinor under Lorentz transformations
\begin{equation}
\chi_{C\, \mu}(x)
= \sum_{a,b,c=1,2,3}\varepsilon^{a b c} \left ( u^{aT}(x)\, C \gamma_\mu \, u^b(x) \right ) u^c(x) \, ,
\end{equation}
where $u^a(x)$ represents the up quark field operator with color index $a$ acting at space-time
coordinate $x$.  Thus, the bare state $|\, \Delta_{C} \, \rangle = \overline \chi_{C\,\mu}(0)\, |\,
\Omega\, \rangle$.  As such, it excites a superposition of QCD energy eigenstates governed by the
smearing extent of $\chi_{C\,\mu}$.  Our hypothesis is that this is the realization of the bare
$\Delta^{++}$ in the Hamiltonian model.

The first step in evaluating $\lambda^{\alpha}_C$ of Eq.~(\ref{eq:lqcd-pv}), and thus
$p^V_{N^*}(E_\alpha,L)$, is to access the spectrum of eigenstates, $|
\Psi^V_{E_\alpha}\rangle$.
This is done via the variational or correlation matrix method
\cite{Michael:1985ne,Luscher:1990ck,McNeile:2000xx,Mahbub:2010rm,Edwards:2011jj,Lang:2012db,Mahbub:2013ala,Roberts:2013ipa,Roberts:2013oea,Alexandrou:2013fsu,Kiratidis:2015vpa}.
The approach involves a matrix of parity-projected correlation functions. In the rest frame of the
state ($\vec p = \vec 0$) the correlation matrix is
\begin{align}
G_{ij}(t,\vec 0) &= \sum_{\vec x} {\rm Tr}_{\rm sp} \left \{ \Gamma_{\pm} \,
  \langle\,\Omega\,\vert\, \chi_{i}(x)\, \overline\chi_{j}(0)\, \vert\,\Omega\,\rangle \right \} \, .
\label{eq:lqcd-gij}
\end{align}
Here, an interpolating field $\overline \chi_{j}(0)$, having the quantum numbers of the considered
state, acts on the QCD vacuum $\vert\,\Omega\,\rangle$ and excites a superposition of finite-volume
energy eigenstates.  The interpolator $\overline \chi_{j}(0)$ is an arbitrary operator,
constrained only by the quantum numbers.  It may be a local operator or a non-local operator
designed to provide overlap with the multi-particle scattering states of the resonance channel.  For
example, operators in which the momentum of each particle in the multi-particle state is specified
are particularly good at exciting these states from the vacuum \cite{Lang:2012db,Lang:2016hnn}.

Appendix A outlines the complete details for calculating $\lambda^{\alpha}_C$ using the correlation
matrix of Eq.~(\ref{eq:lqcd-gij}) and the bare-state definition of Eq.~(\ref{eq:lqcd-pv-1}) for
$|\, N^*_{C} \, \rangle$.  The final result is
\begin{eqnarray}
\left ( p^V_{N_{C}^*}(E_\alpha,L) \right )^{1/2} = \, \lambda^{\alpha}_C \,
= z^{\alpha} \, \frac{G_{Cj}(t)\, u^{\alpha}_{j}}{v^{\alpha}_{i}\, G_{ij}(t)\, u^{\alpha}_{j}}  \, .
\end{eqnarray}
Here the $u^{\alpha}_{i}$ ($v^{\alpha}_{i})$ are the coefficients of the interpolating fields
$\bar{\chi}_i$ ($\chi_i$) forming the optimized interpolating fields $\bar{\phi}^{\alpha} = \sum
u_{i}^{\alpha}\, \overline{\chi}_i$\ \ ($\phi^{\alpha} = \sum v_{i}^{\alpha}\, \chi_i$), designed
to isolate a single energy eigenstate, $\alpha$.  These coefficients are obtained by solving the
generalized eigenvalue problem.  The coefficients $z^{\alpha}$ are the corresponding coupling
strengths between the eigenstate $|\Psi^V_{E_\alpha} \rangle$ and $\bar{\phi}^\alpha\,
|\Omega\rangle$.  In Appendix A, we provide a complete example for nucleon case.

Finally, the averaging and normalization of Eqs.~(\ref{eq:pe-ave}) and (\ref{eq:pe-norm})
respectively provide the final relations for the calculation of the energy-averaged probability
$P^V_{N^*}(E_\alpha,L)$ from $p^V_{N_{C}^*}(E_\alpha,L)$.

In summary, a determination of $P^V_{N^*}(E_\alpha,L)$ in LQCD holds the potential to confirm a
long-standing Ansatz for the internal structure of baryon resonances in coupled-channel analyses.
Giving regard to Figs.~\ref{fg:SLPE} and \ref{fg:1b3cPE}, even a volume with $L = 5$ fm should be
sufficient to disclose a peak in the case of the $P_{33}$ and $P_{11}$ resonances.  We strongly
encourage LQCD groups to calculate $p^V_{N_{C}^*}(E_\alpha,L)$ in future simulations.

\section{Summary and Future Development}

We have investigated the finite-volume Hamiltonian method by
using the meson-exchange model of $\pi N$ reactions within which
bare states are introduced to
parametrize the intrinsic excitations of the nucleon.
In addition to further examining
the differences between the finite-volume Hamiltonian method
and the  L\"uscher formalism,
an approach has been developed to relate
the internal structure of nucleon resonances extracted from the $\pi N$ reaction data to
lattice QCD (LQCD) calculations.

We first showed that the resonance pole positions can be related to the probability $P_{N^*}(E)$ of
finding the bare state in the $\pi N$ scattering states in infinite volume.  We then demonstrated
that the probability, $P_{N^*}^V(E,L)$, of finding the same bare state in the eigenstates of the
underlying Hamiltonian in finite volume approaches $P_{N^*}(E)$ as the volume increases.  Our
findings open the possibility of using $P_{N^*}^V(E,L)$ to examine whether the internal structure
of nucleon resonances extracted from the $\pi N$ reaction data within dynamical models are
consistent with similar measures in LQCD.

We have also discussed possible LQCD calculations of $P_{N^*}^V(E,L)$ under the hypothesis that the
bare states of the dynamical reaction model can be identified with spatially-smeared three-quark
operators acting on the nontrivial vacuum of QCD.  It will be interesting to explore the
results of LQCD calculations of $P_{N^*}^V(E,L)$.

\begin{acknowledgements}
We would like to thank Zhan-Wei Liu, James Zanotti, and Ross Young for the useful discussions.
This work was supported by the U.S. Department of Energy,
Office of Science, Office of Nuclear Physics, Contract
No. DE-AC02-06CH11357.  This research used resources of the National
Energy Research Scientific Computing Center, which is supported by the
Office of Science of the U.S. Department of Energy under Contract
No. DE-AC02-05CH11231, and resources provided on Blues and/or Fusion,
high-performance computing cluster operated by the Laboratory
Computing Resource Center at Argonne National Laboratory.  It was also
supported by the Australian Research Council through the ARC Centre of
Excellence for Particle Physics at the Terascale (CE110001104) and by
grants FL0992247 (AWT), DP151103101 (AWT) and DP150103164 (DBL).
Support from the CNPq (Brasil) through grants 313800/2014-6 and
400826/2014-3 (AWT) are gratefully acknowledged.
It was also supported by the Japan Society for the Promotion of Science (JSPS) KAKENHI Grant No.
JP25800149 (HK).
\end{acknowledgements}

%
\appendix

\section{LQCD calculations of $P_{N^*}^V(E,L)$}

\subsection{Implementation of the Three-Quark Core}

In this Appendix, we use nucleon as an example to show how to determine the three-quark core
contribution to the $\alpha$'th eigenstate $|\, \lambda^{\alpha}_C \,|^2= |\, \langle\, N_{C}^* \,
\vert \, \Psi^V_{E_\alpha}\, \rangle \,|^2 =|\, \langle\, \Omega\, \vert\, \chi_C\, \vert\, \,
\Psi^V_{E_\alpha} \rangle\,|^2$ as defined in Eq.~(\ref{eq:lqcd-pv}).
In practice, there is only one local three-quark operator transforming as a spinor under Lorentz
transformations that has significant overlap with the ground-state nucleon and its radial
excitations
\begin{equation}
\chi_C(x) = \varepsilon^{a b c} \left ( u^{aT}(x)\, C\, \gamma_5 \, d^b(x) \right ) u^c(x) \, .
\end{equation}
Here the subscript $C$ denotes core, indicating both the preferred spin-flavour construction of the
quark core and a preferred smearing extent.  By examining the overlap of this operator with the
states of the spectrum, one can probe the quark-core content of the states.

On the lattice smearing proceeds in a gauge invariant manner  \cite{Gusken:1989qx} through the map
\begin{align}
\psi_{i}(x,t) &=\sum_{x'} F(x,x')\, \psi_{i-1}(x',t) \, ,
\end{align}
where $\psi$ is a quark spinor and
\begin{align}
F(x,x') &=
{(1-\alpha)}\, \delta_{x,x'} + \frac{\alpha}{6} \sum_{\mu=1}^{3} \left [\, U_{\mu}(x)\, \delta_{x',x+\hat\mu}
+ U_{\mu}^{\dagger}(x-\hat\mu)\, \delta_{x',x-\hat\mu}\, \right ]\, ,
\end{align}
includes the lattice gauge-field links, $U_{\mu}(x) = {\mathcal P} \exp \left ( \int_0^a A_\mu( x +
\lambda\, \hat\mu)\, d\lambda \right )$, to maintain gauge invariance.
The smearing parameter $\alpha$ is typically taken to be 0.7 and the smearing extent is governed by
the number of smearing sweeps, $n_s$.  Commencing with a point source in $\psi_{0}(x,t)$, the
smeared operator is
\begin{align}
\psi_{n_s}(x,t) &=\sum_{x'}F^{\,n_s}(x,x')\, \psi_{0}(x',t).
\end{align}
Typically, $n_s \sim 100$ provides optimal overlap with the ground state, corresponding to an RMS
radius of 8.4 lattice units on a $32^2$ lattice volume or 0.84 fm for lattice spacing $a \sim 0.1$
fm.
As this optimal smearing extent includes influence of the meson cloud, it will be interesting
to explore smaller smearing extents more closely related to the quark core, governed by the presence
of coherent centre domain in the QCD vacuum \cite{Stokes:2013oaa}.  To accommodate the node in the
radial wave function of the bare Roper, a superposition of smeared sources of different widths can
be used \cite{Mahbub:2013ala}.

\subsection{Isolation of Excited States}

Accessing the excited states of the spectrum is done via the
variational method or correlation matrix method
 \cite{Michael:1985ne,Luscher:1990ck,McNeile:2000xx,Mahbub:2010rm,Edwards:2011jj,Lang:2012db,Mahbub:2013ala,Roberts:2013ipa,Roberts:2013oea,Alexandrou:2013fsu,Kiratidis:2015vpa}.
The approach involves a matrix of parity-projected correlation
functions.  In the rest frame of the nucleon ($\vec p = \vec 0$) an
$N\times N$ correlation matrix provides
\begin{align}
G_{ij}(t,\vec 0) &= \sum_{\vec x} {\rm Tr}_{\rm sp} \left \{ \Gamma_{\pm} \,
  \langle\,\Omega\,\vert\, \chi_{i}(x)\, \overline\chi_{j}(0)\, \vert\,\Omega\,\rangle \right \} \, .
\end{align}
Here, interpolating field $\overline\chi_{j}(0)$, having the quantum
numbers of the nucleon, acts on the QCD vacuum $\vert\,\Omega\,\rangle$
and excites a superposition of finite-volume energy eigenstates.
These states are annihilated back to the vacuum at space-time $x$.
Summing over all $\vec x$ projects zero momentum and
taking the trace with $\Gamma_{\pm} = \frac{1}{2}(\gamma_0 \pm 1)$
projects positive/negative parity states.
Upon inserting a complete set of intermediate energy eigenstates,
$|\Psi^V_{E_\alpha}\rangle$, with momentum $\vec{p'}$ and spin $s$
\begin{align}
\sum_{\alpha,\, \vec {p'},\, s}\, \vert{\Psi^V_{E_\alpha},\, \vec {p'},\, s}\rangle \,
\langle{\Psi^V_{E_\alpha},\, \vec {p'},\, s}\vert &= I \, ,
\label{eq:completeness}
\end{align}
where $\alpha$ can include multi-particle states, and
using the space-time translation operator
\begin{align}
\chi_{i}(x) &= e^{iP\cdot x}\, \chi_{i}(0)\, e^{-iP\cdot x},
\end{align}
one obtains
\begin{align}
{G_{ij}(t,\vec 0)}
&= \sum_{\alpha}\, \sum_{s}\, {\rm Tr}_{\rm sp} \left \{
\Gamma_{\pm}\, \langle
{\,\Omega\,}\vert\, \chi_{i}(0)\, \vert{\Psi^V_{E_\alpha},\, {\vec 0},\, s}\rangle\,
\langle{\Psi^V_{E_\alpha},\, {\vec 0},\, s}\vert \, \bar\chi_{j}(0)\, \vert
       {\,\Omega\,} \rangle \right \} \, e^{-E_{\alpha}t} \, ,
\label{eq:EucleadianCorrelationFunction}
\end{align}
in Euclidean time. Recalling $E_\alpha$ is the energy of the eigenstate $|\Psi^{V}_{E_\alpha}\rangle$ at rest, i.e., $m_\alpha$.

Focusing on the positive-parity sector of interest herein, the overlap
of the interpolators $\chi_{i}(0)$ with state $\vert{\Psi^V_{E_\alpha},\, {\vec
    0},\, s}\rangle$ is described in terms of the Dirac spinor for
state $\Psi^V_{E_\alpha}$, $u^\alpha({\vec 0},\,s)$, as
\begin{align}
\langle{\,\Omega\,}\vert\, \chi_i(0)\, \vert \Psi^V_{E_\alpha},\, \vec 0,\, s\rangle
&= \lambda^\alpha_i \, u^\alpha({\vec 0},\,s) \, ,
\end{align}
and
\begin{align}
\langle \Psi^V_{E_\alpha},\, \vec 0,\, s\vert\, \bar\chi_j(0)\, \vert {\,\Omega\,} \rangle
&=\bar\lambda^\alpha_j \, \bar u^\alpha({\vec 0},s) \, .
\end{align}
Here, $\lambda_{i}^{\alpha}$ and $\bar\lambda_{j}^{\alpha}$ are the
couplings of interpolators $\chi_{i}$ and  $\bar\chi_{j}$ at the sink
and source respectively to eigenstates $\alpha=0, \cdots, (N-1)$.
Recalling
\begin{align}
\sum_{s} u^{\alpha}(\vec p,s)\, {\bar u}^{\alpha}(\vec p,s)
  &=\frac{\gamma \cdot p + m_\alpha}{2\,\sqrt{m^2_\alpha + \vec{p}^2}} \, ,
\end{align}
and taking the spinor trace
\begin{align}
G_{ij}(t, \vec 0) &= \sum_{\alpha=0}^{N-1}\, \lambda_{i}^{\alpha}\, \bar\lambda_{j}^{\alpha} \,
  e^{-m_{\alpha}\, t} \, .
\end{align}

The interpolating fields provide an $N$-dimensional basis upon which
to describe the $N$ lowest-lying states.  Using this basis, we seek
linear combinations which isolate each state, $\alpha$
\begin{align}
{\bar\phi}^{\alpha} &=\sum_{i=1}^{N}\, u_{i}^{\alpha}\, {\bar\chi}_{i},
\qquad
{\phi}^{\alpha} =\sum_{i=1}^{N}\, v_{i}^{\alpha}\, {\chi}_{i},
\end{align}
such that,
\begin{align}
\langle{\Psi^V_{E_\beta},\, \vec p,\, s}\vert\, {\bar\phi}^{\alpha}\,
\vert\,\Omega\,\rangle &= \delta_{\alpha\beta}\, {\bar{z}}^{\alpha}\,
\bar{u}^\alpha(\vec p,\, s)\, ,
\quad\mbox{and}\quad
\langle\,\Omega\,\vert\, {\phi}^{\alpha}\, \vert \Psi^V_{E_\beta},\, \vec p,\, s\rangle =
\delta_{\alpha\beta}\, {z}^{\alpha}\, u^\alpha(\vec p,\, s)\, .
\end{align}
Here $z^{\alpha}$ and ${\bar{z}}^{\alpha}$ are the coupling strengths
of $\phi^{\alpha}$ and ${\bar\phi}^{\alpha}$ to the state $\vert
\Psi^V_{E_\alpha},\, \vec p,\, s\rangle$.

By multiplying the correlation matrix $G_{ij}(t)$ by
$u_{j}^{\alpha}$ and summing over repeated indices, one obtains
\begin{subequations}
\begin{align}
G_{ij}(t,\vec 0)\, u_{j}^{\alpha}
&= \sum_{\vec x} {\rm Tr}_{\rm sp} \left \{ \Gamma_{\pm} \,
  \langle\,\Omega\,\vert\, \chi_{i}(x)\, \overline\chi_{j}(0)\, u_{j}^{\alpha} \, \vert\,\Omega\,\rangle \right \} \, ,
\label{eq:rightEVa}\\
&= \sum_{\vec x} {\rm Tr}_{\rm sp} \left \{ \Gamma_{\pm} \,
  \langle\,\Omega\,\vert\, \chi_{i}(x)\, \overline\phi_{j}(0) \, \vert\,\Omega\,\rangle \right \} \, ,
\label{eq:rightEVb}\\
&= \lambda_{i}^{\alpha}\, \bar{z}^{\alpha}\, e^{-m_{\alpha}\, t}\, .
\label{eq:rightEVc}
\end{align}
\end{subequations}
illustrating the time dependence is described by the mass of the eigenstate energy.  Since the $t$
dependence is described by the exponential term alone, a recurrence relation at times $t$ and
$t+\Delta t$ constructed
\begin{align}
G_{ij}(t+\Delta t)\, u_{j}^{\alpha} &= e^{-m_{\alpha}\, \Delta t}\, G_{ij}(t)\, u_{j}^{\alpha} \, .
\end{align}
This generalized eigenvalue equation can be solved for eigenvectors $\mathbf{u}^\alpha$ with
eigenvalues $\exp(-m_{\alpha}\, \Delta t)$.  Similarly
\begin{align}
v_{i}^{\alpha}\, G_{ij}(t+\Delta t) &= e^{-m_{\alpha}\, \Delta t}\, v_{i}^{\alpha}\, G_{ij}(t) \, .
\end{align}
defines the left eigenvector $\mathbf{v}^\alpha$.
With the eigenvectors normalized in the usual manner $\mathbf{u}^{\dagger \alpha} \mathbf{u}^\alpha
= \mathbf{v}^{\dagger \alpha} \mathbf{v}^\alpha = 1$, the coupling strengths $z^\alpha$ and $\bar
z^\alpha$ are defined.

The eigenvectors $u_{j}^{\alpha}$ and $v_{i}^{\alpha}$ can then be used to create the
projected correlator
\begin{align}
v_{i}^{\alpha}\, G_{ij}(t)\, u_{j}^{\beta} = \delta^{\alpha\beta}\, z^{\alpha}\,
{\bar{z}}^{\beta}\, e^{-m_{\alpha}\, t}\, .
\label{eq:projCorr}
\end{align}


In the ensemble average the correlation matrix is symmetric and therefore one usually works with
the improved unbiased estimator $(\, G_{ij}(t) + G_{ji}(t)\, )/2$.  Because the QCD action is the
same for link ensembles $\{ U_\mu(x) \}$ and $\{ U^*_\mu(x) \}$ one can show that the two point
correlation functions of the correlation matrix can be made to be perfectly real
 \cite{8Draper:1988xv,Leinweber:1990dv,Boinepalli:2006xd}.  Averaging the link ensembles $\{ U_\mu(x)
\}$ and $\{ U^*_\mu(x) \}$ and ensuring $G$ is symmetric for each configuration ensures the
coupling strengths are real and $\bar\lambda^\alpha_i = \lambda^\alpha_i$ and $\bar z^\alpha =
z^\alpha$.

\subsection{Determining the strength of the core}

We are now in a position to determine the overlap of lattice-QCD energy eigenstate $\vert \Psi^V_{E_\alpha}
\rangle$ with the three-quark core, $\langle\, N_C^*\, \vert\, \Psi^V_{E_\alpha}\, \rangle$.  Using
the projected correlator of Eq.~(\ref{eq:projCorr}) the overlap of the eigenstate interpolators
$\phi^\alpha$ and $\bar\phi^\alpha$ is determined by a linear fit to the logarithm of the projected
correlator
\begin{align}
\log\left ( v_{i}^{\alpha}\, G_{ij}(t)\, u_{j}^{\alpha} \right) =   2\, \log \left( z^{\alpha}
\right ) - m_{\alpha}\, t\, .
\label{eq:logProjCorr}
\end{align}
The core contribution can be isolated via Eqs.~(\ref{eq:rightEVb}) and (\ref{eq:rightEVc}).
Replacing $\langle\,\Omega\,\vert\, \chi_i(x)$ by the core contribution
$\langle\,\Omega\,\vert\, \chi_C(x) =\langle\, N_C^* \,\vert$,
the core contribution to eigenstate  $\vert\, \Psi^V_{E_\alpha}
\rangle$, $\langle\, N_C^* \,\vert\, \Psi^V_{E_\alpha} \rangle =
\lambda_C^\alpha$ is obtained via
\begin{align}
\log\left ( G_{Cj}(t)\, u_{j}^{\alpha} \right) = \log\left( \lambda_C^{\alpha} \right ) + \log\left( z^{\alpha} \right )
- m_{\alpha}\, t\, .
\label{eq:logCCorr}
\end{align}
where
\begin{align}
G_{Cj}(t)
&= \sum_{\vec x} {\rm Tr}_{\rm sp} \left \{ \Gamma_{\pm} \,
  \langle\,\Omega\,\vert\, \chi_{C}(x)\, \overline\chi_{j}(0)\, \vert\,\Omega\,\rangle \right \} \, .
\label{eq:Gcj}
\end{align}
Here the time dependence can be eliminated through a ratio such that
\begin{align}
\lambda_C^{\alpha} =
z^\alpha \, \frac{G_{Cj}(t)\, u_{j}^{\alpha}}{v_{i}^{\alpha}\, G_{ij}(t)\, u_{j}^{\alpha}}  \, .
\label{eq:CorrRatio}
\end{align}

\section{The Study of  fluctuation of $p^V_{\Delta}(E)$}

For understanding the fluctuation of $p_\Delta^V(E,L)$ shown in Fig.4, we consider an exactly
soluble model which has one bare state and one channel ($1b1c$) to describe the $P_{33}$ $\pi N$
scattering.  The Hamiltonian of this $1b1c$ model only has a bare $\Delta \rightarrow \pi N$
interaction:
\begin{eqnarray}
\Gamma(k)\equiv \langle k| g | \Delta \rangle &=&  \frac{g}{\sqrt{m_\pi}}\frac{k}{\sqrt{m^2_\pi+k^2}}\frac{1}{(1+(k/\Lambda)^2)^2}\frac{1}{\sqrt{1+(k/\Lambda)^2}},
\end{eqnarray}
where $g$ and $\Lambda$ are the bare coupling and cut off, $m_\pi$ is the mass of pion.  As shown
in Fig.~\ref{fg:phasea}, the $P_{33}$ phase shifts generated from the SL model can be reproduced by
choosing : $g=0.30390$, $\Lambda=656.60$ MeV, and $m_0=1265.04$ MeV for the mass of the bare
$\Delta$.


\begin{figure}[tbp] \vspace{-0.cm}
\begin{center}
\includegraphics[width=0.6\columnwidth]{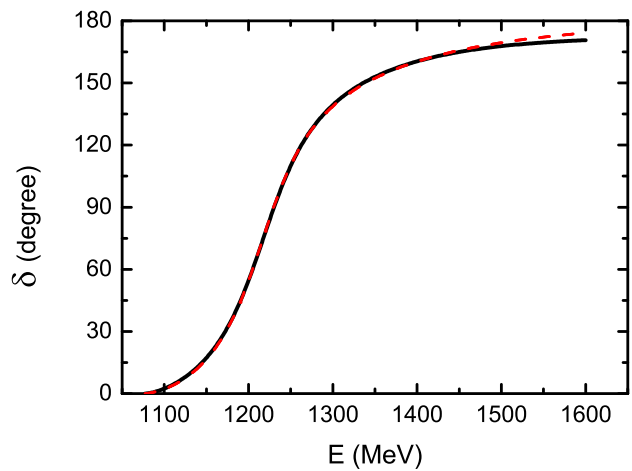}
\caption{The black solid and red dashed lines are calculated from SL model and 1b1c model, respectively.}
\label{fg:phasea}
\end{center}
\end{figure}

\begin{figure}[tbp] \vspace{-0.cm}
\begin{center}
\includegraphics[width=0.6\columnwidth]{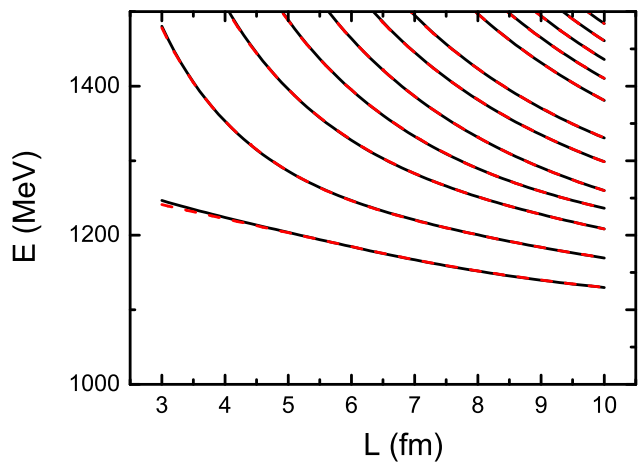}
\caption{The spectrum of $\pi N$ in the finite volume. The black solid and red dashed lines are calculated from SL model and 1b1c model, respectively.}
\label{fg:specta}
\end{center}
\end{figure}

Within this 1b1c model, we need
to find the eigenvalues $E_i$ and eigenstate $|\Psi^V(E_i)>$ from
the Hamiltonian matrix of the following form:

\begin{eqnarray}
[H]_{N+1}&=&\left( \begin{array}{ccccc}
m_0                     & g^V(k_0)                          & g^V(k_1)                    & \cdots                          & g^V(k_{N-1})  \\
 g^V(k_0)            & E_\pi(k_0)+E_N(k_0)      & 0                                  & \cdots                          & 0                      \\
 g^V(k_1)            & 0                                      &  E_\pi(k_1)+E_N(k_1) & \cdots                          & 0                       \\
\vdots                  & \vdots                             & \vdots                          & \ddots                         & \vdots                \\
 g^V(k_{N-1})     & 0                                     & 0                                   & \cdots                         & E_\pi(k_{N-1})+E_N(k_{N-1}), \\
\end{array} \right) \nonumber
\end{eqnarray}
where $k_n=\sqrt{n}2\pi/L$ for integers $n=0,1,2\cdot\cdot\cdot$, as
 specified by the quantization condition in finite volume
with size $L$, and
\begin{eqnarray}
g^{V}(k_n)&=&\sqrt{\frac{C_3(n)}{4\pi}}\left(\frac{2\pi}{L}\right)^{3/2}
\Gamma(k_n).
\label{eq:vfin2-1}
\end{eqnarray}
Here $C_3(n)$ is the number of degenerate states with the same magnitude,
$k_n=|\vec{k}_n|$.

With the simple matrix $[H]_{N+1}$ given above, it is easy to see that Eq.~(24) for finding
 the eigenvalues become
\begin{eqnarray}
 E_i - m_0 -\Sigma^V(E_i, L)=0.\label{eq:equation}
\end{eqnarray}
where the self-energy is
\begin{eqnarray}
\Sigma^V(E,L) \equiv \sum_{n}\left(\frac{2\pi}{L}\right)^{3}\frac{C_3(n)}{4\pi} \frac{\Gamma( k_n)\Gamma^{*}(k_n)}{E-E_{\pi}(k_n)-E_{N}(k_n)} \label{eq:Sigma}
\end{eqnarray}
The solutions of Eq.~(\ref{eq:equation}) reproduce the spectrum of the SL model, as shown
in Fig.~\ref{fg:specta}.
The eigenstate $| \Psi^V(E_i)\rangle$  can also be solved exactly:
\begin{eqnarray}
| \Psi^V(E_i)\rangle &=& \frac{1}{\sqrt{Z(E_i, L)}}\left[|\Delta\rangle + \sum_{n} \sqrt{\frac{C(n)}{4\pi}} \left(
\frac{2\pi}{L}\right)^{\frac{3}{2}}
\frac{\Gamma(k_{n}) }{E_{i}-E_{\pi}(k_{n})-E_{N}(k_n)} |k_n\rangle\right],\label{eq:eigenstate}
\end{eqnarray}
where the normalization constant is
\begin{eqnarray}
Z(E_i, L) &=& 1 +  \sum_{n}\left(\frac{2\pi}{L}\right)^{3} \frac{C(n)}{4\pi}\frac{\Gamma^{*}(k_n) \Gamma(k_n) }{(E_{i}-E_{\pi}(k_n)-E_{N}(k_n))^2} .\label{eq:zi}
\end{eqnarray}
From Eqs.~(\ref{eq:Sigma}) and (\ref{eq:zi}), we have the following relation
\begin{eqnarray}
Z(E_i,L)= \left . 1-\frac{\partial\Sigma^V(E,L)}{\partial E} \right |_{E=E_i}
\end{eqnarray}
From Eqs.~(\ref{eq:eigenstate}) and (\ref{eq:zi}), we then have
\begin{eqnarray}
p_\Delta^V(E_i,L)= \left | \langle \Delta | \Psi^V(E_i)\, \rangle \right |^2 &=& \frac{1}{Z(E_i, L)}\nonumber \\
&=&\frac{1}{\left . 1-\frac{\partial\Sigma^V(E,L)}{\partial E}\right |_{E=E_i}}
\label{eq:pvea}
\end{eqnarray}
The resulting $p_\Delta^V(E_i,L)$ for various volume sizes $L$ are similar to that shown in Fig.4
for the SL model.  Here we only show the result of $L=10$ in the left side of Fig.~\ref{fg:points}.

Obviously, $p_\Delta^V(E_i, L)$ also shows fluctuations within this exactly soluble $1b1c$ model.
To understand this, we show $\Sigma^V(E_i, L)$ (black solid curves) and $E-m_0$ (red dashed line)
in the right-hand panel of Fig.~\ref{fg:points}.  From Eq.~(\ref{eq:equation}), it is obvious that
the $i$-th solid green dot in the right side is the eigenvalue $E_i$ for each $p_\Delta^V(E_i,L)$ shown in the
left side of the figure.  From the expression Eq.~(\ref{eq:zi}), we see that when an eigenvalue
$E_i$ is close to any of the energy grid points, $\epsilon(k_n) \equiv E_{\pi}(k_n)+E_{N}(k_n)$,
the normalization constant $Z(E_i,L)\rightarrow \infty$ and hence $p_\Delta^V(E_i,L)$, as defined in
Eq.~(\ref{eq:pvea}), becomes negligible. It is also clear that if $E_i$ is farther away from the
energy grid points, $Z(E_i,L)$ will be smaller and hence $p_\Delta^V(E_i,L)$ will be larger. We can see this
clearly by comparing the values of $p_\Delta^V(E_i,L)$ (black dots in the left side) for the 1-st to 4-th
eigenvalues and the distances between the corresponding green dots and the nearest energy grid
points in the right side.  Similar comparisons also explain the fluctuation between 4-th and 8th
eigenvalues.  The peak at the 7-th eigenvalue in the left side can be understood as follows.  The gap
between two grid energies near the 7-th eigenvalue is much larger than the distances between any
other two energy grids, since there is no integer vector which has a length equal to $\sqrt{7}$.
As a result, the self energy $\Sigma^V(E,L)$ has a smaller slope near the 7-th eigenvalue and hence $-
\left . \frac{\partial \Sigma^V(E, L)}{\partial E} \right |_{E=E_7}$ is smaller than those of the 6-th and 8-th
eigenvalues.  This can be seen in Table \ref{tab:pve}. The fluctuations in other areas can also be
understood from Eq.~(\ref{eq:pvea}) and the values listed in Table \ref{tab:pve}.

In summary, the fluctuation in $p_\Delta^V(E_i,L)$ is the mathematical consequence of the special
property of the lattice momenta specified by the quantization condition in finite volume. While
this can be proved unambiguously only within this exactly soluble $1b1c$ model, it does provide an
explanation for the fluctuations seen in Fig.4 for the more realistic SL model.

\begin{figure}[tbp] \vspace{-0.cm}
\begin{center}
\includegraphics[width=0.49\columnwidth]{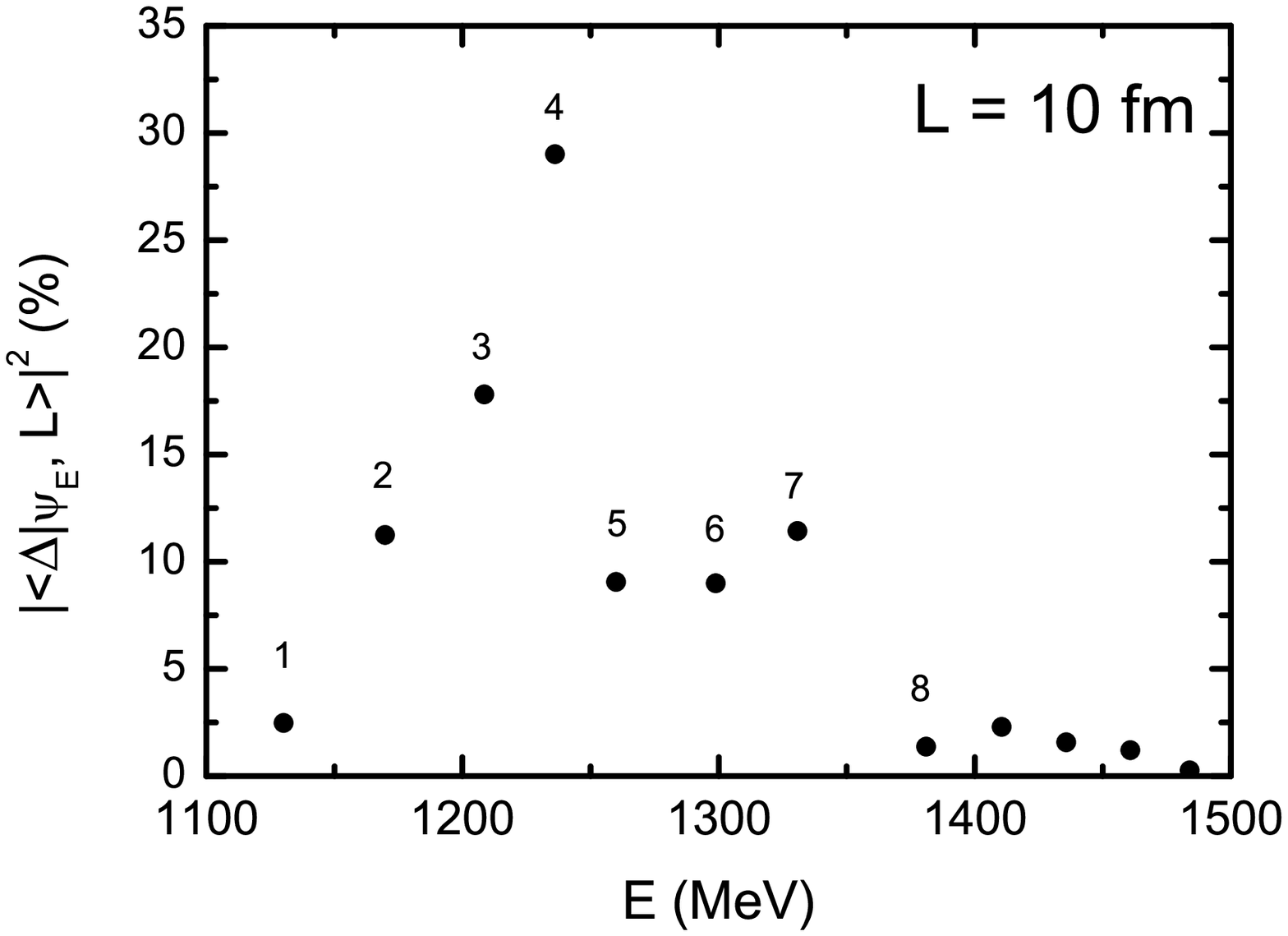}
\includegraphics[width=0.49\columnwidth]{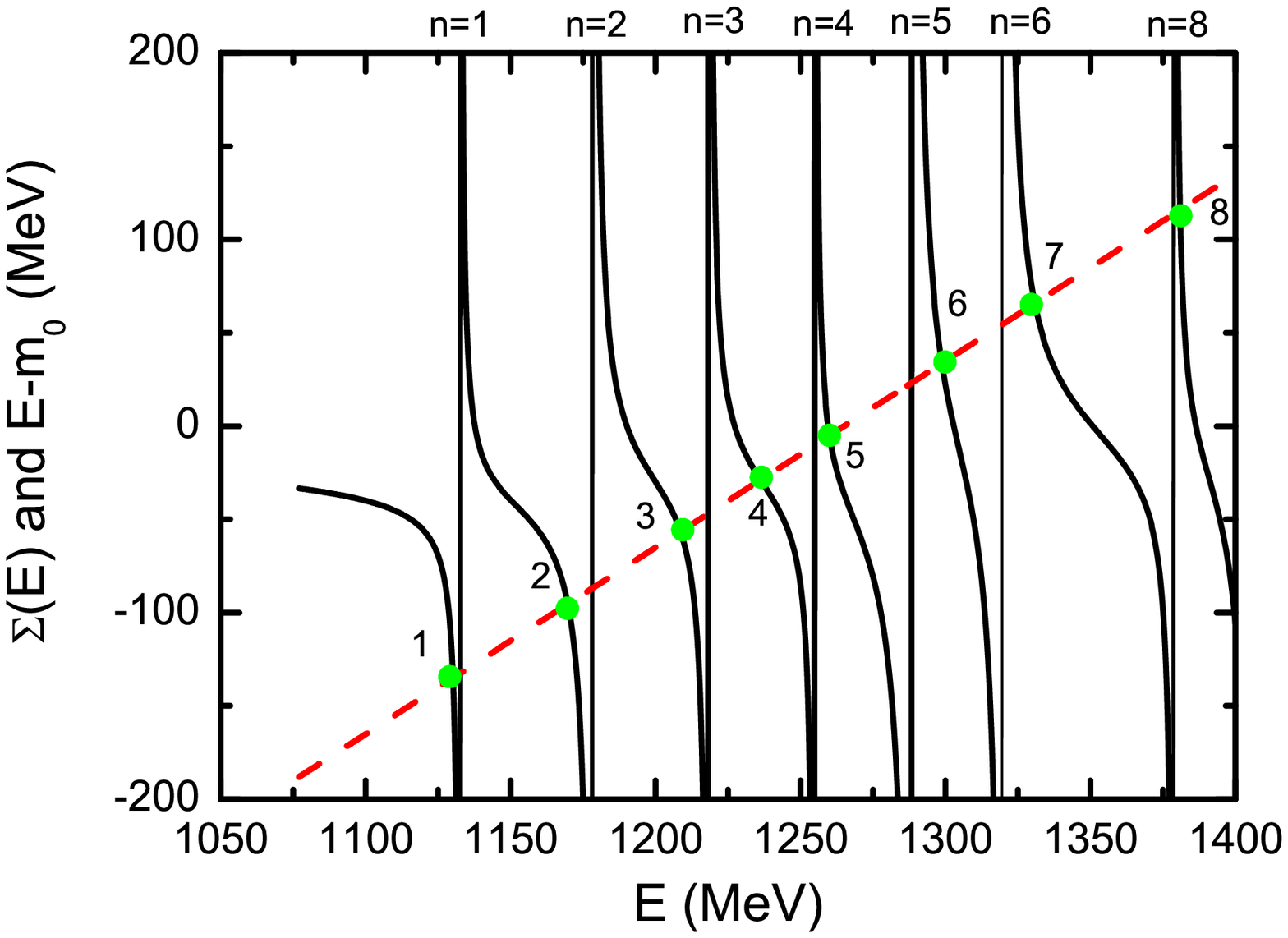}
\caption{The left plane is $| \langle \Delta | \Psi^V(E_i) \rangle |^2$ vs $E_i$ at $L=10$ fm. In
  the right panel, the black solid lines and red dashed line are the functions $\Sigma^V(E)$ and
  $E-m_0$ respectively as a function of energy $E$. The green solid points are the crossing points
  of the black and red lines, corresponding to the eigenvalues of the Hamiltonian
  .}
\label{fg:points}
\end{center}
\end{figure}

\begin{table}[ht]
\begin{center}\caption{ The value of $-\left . \frac{\partial \Sigma^V(E)}{\partial
      E}\right |_{E=E_i}$ in the $1b1c$ model.}
\begin{ruledtabular}
\begin{tabular}{ccc}
    $i$      &   $E_i$ (MeV)       &   $-\left . \frac{\partial \Sigma^V(E)}{\partial E}\right |_{E=E_i}$   \\
\hline
    $1$      &   1130.1            &   38.8                                                        \\
    $2$      &   1169.7            &   7.88                                                        \\
    $3$      &   1208.5            &   4.51                                                        \\
    $4$      &   1236.2            &   2.45                                                        \\
    $5$      &   1260.0            &   10.1                                                        \\
    $6$      &   1298.8            &   10.2                                                        \\
    $7$      &   1330.8            &   7.70                                                        \\
    $8$      &   1381.0            &   75.9                                                        \\
\end{tabular}  \label{tab:pve}
\end{ruledtabular}
\end{center}
\end{table}

\end{document}